\documentclass[
 reprint,
  amsmath,amssymb,
  aps,superscriptaddress,
 prb,nobibnotes
 ]{revtex4-1}
 \usepackage{graphicx}
 \usepackage{dcolumn}
 \usepackage{float}
 \usepackage{placeins}
 \usepackage[mathscr]{euscript}
 \usepackage{xcolor}
 \usepackage[linkbordercolor=green]{hyperref}
 \usepackage{tikz}
 \hypersetup{
     colorlinks=true,
     linkcolor=red,
     filecolor=magenta,
     urlcolor=cyan,
     citecolor = red,
 }
\usepackage{rotating}
\usepackage{enumitem}

 \usepackage{amsmath}
 
\usepackage{multirow}
\usepackage{bm}

\usepackage[normalem]{ulem}

\usepackage{capt-of}

\usepackage{lineno}

\newcommand{\rr}{\bm{r}}
\newcommand{\hr}{\hat{\bm{r}}}

\newcommand{\uu}{\bm{u}}
\newcommand{\hu}{\hat{\bm{u}}}
\newcommand{\vv}{\bm{v}}
\newcommand{\pp}{\bm{p}}
\newcommand{\FF}{\bm{F}}

\newcommand{\II}{\bm{I}}
\newcommand{\jj}{\bm{j}}
\newcommand{\ee}{\bm{e}}
\newcommand{\NN}{\bm{N}}
\newcommand{\PP}{\bm{P}}
\newcommand{\xx}{\bm{\xi}}
\newcommand{\ww}{\bm{\omega}}

\newcommand{\ttt}{\bm{\tau}}
\newcommand{\one}{\bm{1}}
\newcommand{\tunit}{t_{_0}}

\newcommand{\ud}{\text{d}}
\newcommand{\ue}{\text{e}}
\newcommand{\kbT}{k_{\text{B}}T}

\newcommand{\rev}[1]{\textcolor{black}{#1}}

 \date{January 2026}

\begin{document}

 \title{A general model for frictional contacts in colloidal systems}
 
\author{Kay Hofmann}
\email{khofmann@uni-mainz.de}
\affiliation{Institute for Physics, Johannes Gutenberg-Universität Mainz, 55099 Mainz, Germany}

\author{Kay-Robert Dormann}
\email{kay-robert.dormann@pkm.tu-darmstadt.de}
\affiliation{Institute for Condensed Matter Physics, Department of Physics, Technische Universit\"at Darmstadt, 64289 Darmstadt, Germany}

\author{Benno Liebchen}
\email{benno.liebchen@pkm.tu-darmstadt.de}
\affiliation{Institute for Condensed Matter Physics, Department of Physics, Technische Universit\"at Darmstadt, 64289 Darmstadt, Germany}

\author{Friederike Schmid}
\email{friederike.schmid@uni-mainz.de}
\affiliation{Institute for Physics, Johannes Gutenberg-Universität Mainz, 55099 Mainz, Germany}


\begin{abstract}
In simulations of colloidal matter, frictional contacts between
particles are often neglected. For spherical colloids, such an
approximation can be problematic, since frictional contacts couple
translational and rotational degrees of freedom, which may affect the
collective behavior of, e.g., colloids under shear and chiral active
matter. Deterministic models for frictional contacts have been
proposed in the granular matter community.  On the colloidal scale,
however, thermal fluctuations are important and should be included in
a thermodynamically consistent manner.  Here, we derive the correct
fluctuation-dissipation relation for linear and nonlinear
instantaneous frictional contact interactions. Among other, this
generates a new generalized class of dissipative particle dynamics
(DPD) thermostats with rotation-translation coupling. We demonstrate
effects of frictional contact interactions using the examples of
Poiseuille flow and motility induced phase separation in active
Langevin particles.
\end{abstract}

\maketitle 




\section{Introduction}

Friction has been studied for centuries. Early important observations
of frictional forces acting on macroscopic bodies on surfaces have
been made by da Vinci and Amontons \cite{bowden1973friction}. In the
18th century, Coulomb discovered his famous friction law
\cite{popova2015research}.  Later, the role of frictional contacts
between particles in many particle systems has also been intensively
explored in granular matter. Here, friction induces a significant
coupling between translational and rotational degrees of freedom of
the particles. Commonly employed standard models describing frictional
contacts in granular matter have been discussed in Refs.~
\cite{cundall1979discrete,elperin1997comparing,herrmann1998modeling,luding2005anisotropy,silbert2001granular, silbert2010jamming}.
Characteristically, these models do not account for thermal
fluctuations, which are negligibly weak for granular particles.  A
first model to describe dry friction in Brownian particles has been
developed by de Gennes in 2005 \cite{gennes2005brownian}. This paper
also proposes an experimental realization based on granular particles
in randomly vibrating plates. Such situations, which involve both
frictional contacts and imposed Brownian motion, have subsequently
been studied  both theoretically and experimentally~\cite{touchette2010brownian,baule2012singular, 1FDT_1972,semeraro2023diffusion}
and very recently also in (granular) active particles
\cite{antonov2024inertial}. 

At the colloidal scale, frictional contacts have long been neglected.
However, within the past decade, there has been increasing evidence that
frictional contacts can play an important role in colloidal systems, in
particular when they are sheared, leading to large relative velocities
of adjacent colloids.  In particular, the role of frictional contacts
for discontinuous shear thickening in colloidal (and larger) particles
has been prominently discussed.  While discontinuous shear-thickening
has originally been attributed to lubrication flows
\cite{wagner2009shear}, later theoretical
\cite{seto2013discont,wyart2014discontinuous,van2022emergence} and
experimental
\cite{guy2015towards,lin2015hydrodynamic,royer2016rheological,comtet2017pairwise,hsu2018roughness}
works provide evidence, that (nanoscale) repulsive and frictional forces
play a key role for discontinuous shear-thickening. A popular idea is
that at low pressure, the solvent in the gaps between adjacent particles
keeps them apart, whereas at higher pressure, the particles overcome
repulsive forces and frictional contacts become important, ultimately
resulting in shear thickening. While the overall mechanism for the
emergence of discontinuous shear thickening is probably comparatively
involved \cite{morris2020shear}, there is now strong evidence that
frictional contacts play a central role for sheared colloids undergoing
discontinuous shear-thickening. 

Recently, frictional contacts have also been intensively studied in
(colloidal) active matter systems.  Similarly as in sheared systems,
also in active systems the relative velocity between adjacent
particles can be rather high, suggesting that frictional contacts
could play an important role in these systems
\cite{nie2020frictional}.  To describe frictional contacts, Nie et al
\cite{nie2020frictional} have employed a model from the granular
community, whereas Abaurrea-Velasco et al
\cite{abaurrea2020autonomously,narinder2022understanding} have
developed a new model describing overdamped active particles with
frictional contacts. Besides self-propelled active particles, also
active spinners with frictional contacts have been studied
\cite{nguyen2014emergent}.

Importantly, unlike for granular matter, for colloidal matter thermal
fluctuations are important.  Despite this, most models that have been
used to describe frictional contacts in colloidal matter have borrowed
models to describe friction from the granular community
\cite{cundall1979discrete,luding2005anisotropy}, or have developed new
ones \cite{abaurrea2020autonomously,narinder2022understanding} which
have been combined with standard translational and rotational additive
white noise terms, similarly as for particles without frictional
contacts. This may be sufficient to describe common experiments, but
from a fundamental perspective, this approach leads to thermodynamically
inconsistent models that may violate Onsager relations and do not obey
the fluctuation dissipation theorem in the equilibrium limit. This
happens for models that account only for fluctuations that appear as a
counterpart to Stokes drag, but not for the additional fluctuations that
necessarily arise as a counterpart to tangential friction (frictional
contacts) in colloids that are subject to a thermal bath.  As we show in
the present work, thermodynamic inconsistencies due to the consideration
of incomplete thermal fluctuations can (wrongly) lead to the prediction
of different equilibrium temperature values for the rotational and
translational degrees of freedom, which both differ from the bath
temperature, and are accompanied by non-Maxwell-Boltzmann velocity
distributions. In nonequilibrium, for active systems, thermodynamically
inconsistent treatments of frictional contacts can have severe
consequences for the predicted collective behavior.

The central aim of the present work is therefore to develop a
thermodynamically consistent model to describe colloidal matter with
frictional contacts (active and passive). To achieve this, we develop a
model that is closely related to a popular model from the granular
community \cite{herrmann1998modeling} and systematically calculate the
required \emph{multiplicative} noise term.  Using particle-based
simulations, we confirm that the resulting translational and angular
velocity distributions are fully consistent with the predictions from
statistical mechanics and both lead to temperatures that coincide with
the bath temperature.  To exemplarily study the role of thermodynamic
(in)consistencies, we explore two examples for sheared colloids and for
active colloids undergoing motility-induced phase separation. 

Our results could serve as a useful ingredient to model sheared
colloidal suspensions, to help understand discontinuous shear thickening
and to model systems of self-propelled particles as well as active
\cite{soni2019odd, tan2022odd} (and passive \cite{zhao2022odd}) spinners
with frictional contacts. The latter should be of particular interest in
the context of the enormous present interest in odd hydrodynamics
\cite{huang2023odd,fruchart2023odd}.


\section{Results}

\subsection{Theory: Modeling tangential friction interactions}
\label{sec:theory}

\subsubsection{Statement of the problem}

We consider systems of spherical colloidal particles that interact
with each other via conservative forces and surface friction forces.
In addition, the particles may interact with a background medium
and/or experience further driving or active forces.  Specific examples
will be discussed in Section \ref{sec:applications}. Here, we focus on
the modeling of the surface friction forces. 

The friction forces depend on the relative tangential velocities of
the particles at the point of contact, but not necessarily in a linear
manner. If two particles collide, this induces changes of both their
momenta and angular momenta. As mentioned in the introduction,
deterministic models for such contact forces have been discussed for
centuries. On colloidal scales, however, an additional aspect comes
in: Thermal fluctuations become important. Friction forces are
invariably associated with random, stochastic forces, which can not be
ignored on small scales, but must be included in a thermodynamically
consistent manner.  To ensure that passive systems without
nonequilibrium driving forces reach thermal equilibrium, the
frictional and the corresponding stochastic forces must satisfy a
fluctuation-dissipation relation. In other words: the frictional and
stochastic forces must be coupled such that, together, they act as a
thermostat for the system, independent of other thermostats that may
describe, e.g., the coupling to a background medium. 

Our aim is to derive a general structure of thermostats that arise
from tangential friction forces between particles. We will focus on
Markovian forces without memory.  In the granular matter literature,
non-Markovian friction models have also been proposed to mimic
transitions between static and dynamic friction
\cite{elperin1997comparing,herrmann1998modeling}. Such models will not
be considered here.

\subsubsection{Structure of deterministic friction force}

\begin{figure}
    \centering
    \includegraphics[width=0.7\linewidth]{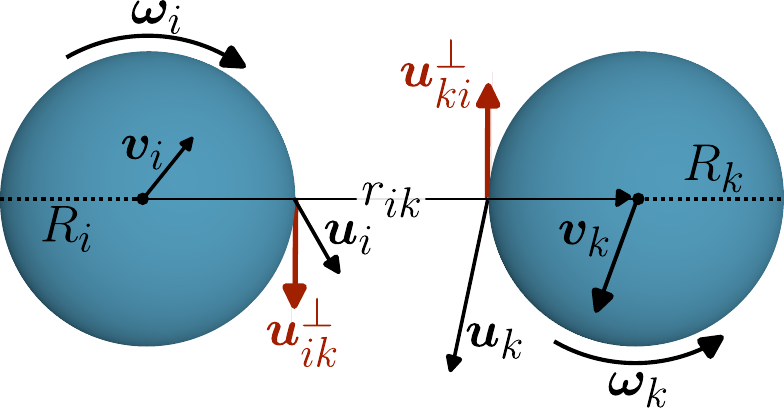}
    \caption{Sketch of a two-particle configuration subject to 
    tangential friction}
    \label{fig:cartoon}
\end{figure}

The general form of deterministic Markovian tangential friction forces
between two spherical particles $i$ and $k$ with radii $R_{i,k}$,
center positions $\rr_{i,k}$, angular velocities $\ww_{i,k}$, and
translational velocities $\vv_{i,k}$, can be constructed as
follows.\cite{elperin1997comparing} If the particles are in contact,
their surface velocities at the contact point are given by
\begin{align}
    \uu_i &= \vv_i  + \ww_i \times \hr_{ik} \: R_i \label{eq:ui} \\
    \uu_k &= \vv_k  - \ww_k \times \hr_{ik} \: R_k, \label{eq:uk}
\end{align}

where we have defined $\hr_{ik} = \rr_{ik}/r_{ik}$ with $\rr_{ik} =
\rr_k - \rr_i$ and $r_{ik} = |\rr_{ik}|$. From these expressions, we
can calculate the relative tangential velocity $\uu_{ik}^\perp$ at the
contact point as the projection of the surface velocity difference,
$\uu_k - \uu_i$, onto the tangent plane at contact (see
Fig.~\ref{fig:cartoon}), i.e., ${\uu_{ik}^\perp = \PP(\hr_{ik}) \:
(\uu_k - \uu_i)}$ with the projection operator
\begin{align}
    \PP(\hr_{ik}) = \one - \hr_{ik}\hr_{ik},
  \label{eq:projection}
\end{align}
giving
  \begin{align}
        \uu_{ik}^\perp  &= \PP(\hr_{ik}) (\vv_k - \vv_i) - (\ww_i R_i + \ww_k R_k) 
   \times \hr_{ik}. \label{eq:uik_perp}
   \end{align}
Assuming that the tangential friction force at the contact point points
in the direction of $\uu_{ik}^\perp$, we model it very generally as
\begin{align}
    \FF_{i}^\text{f,contact} = -\FF_{k}^\text{f,contact} 
   = f(u_{ik}^\perp,r_{ik}) \; \hu_{ik}^\perp
\label{eq:Ffcontact}
\end{align}
with $\hu_{ik}^\perp = \uu_{ik}^\perp/u_{ik}^\perp$, where
$f(u_{ik}^\perp, r_{ik})$ is an arbitrary function of the
relative tangential velocity and the distance between the particles.
The surface force $\FF_{i}^{\text{f,contact}}$ generates a center-of-mass 
force and torque acting on particle $i$,
\begin{equation}
\FF_{ik}^{\text{f}} = \FF_i^{\text{f,contact}}, \qquad
    \ttt_{ik}^{\text{f}} = R_i \: \hr_{ik} \times
\FF_{i}^{\text{f,contact}},
\end{equation}
which is a pair friction force and torque resulting
from the friction with the particle $k$ and can be rewritten as
\begin{equation}
\label{eq:transverse_force}
    \FF_{ik}^{\text{f}} = \frac{f(u_{ik}^\perp,r_{ik})}{u_{ik}^\perp} \:
    \big[ \PP(\hr_{ik}) (\vv_k - \vv_i) 
      + \hr_{ik} \times (\ww_i R_i + \ww_k R_k) \big],
\end{equation}
\begin{equation}
\label{eq:transverse_torque}
\frac{\ttt_{ik}^{\text{f}}}{R_i} = 
   \frac{f(u_{ik}^\perp, r_{ik})}{u_{ik}^\perp} \: 
   \big[ \hr_{ik} \times (\vv_k - \vv_i) 
    - \PP(\hr_{ik})(\ww_i R_i + \ww_k R_k) \big].
\end{equation}
We note that the torques transmitted to the two particles $i$ and $k$
point in the same direction (and are in fact identical for $R_i=R_k$),
whereas the forces are opposed to each other. One can easily
see that the deterministic friction force always reduces
the kinetic energy of the two particles during a collision, 
since
\begin{align}
 \frac{\ud}{\ud t} E_\text{kin} \Big|_\text{f}
 &= \frac{\ud}{\ud t} \frac{1}{2} \Big(
      m_i \vv_i^2 + m_k \vv_k^2
 + \ww_i \II_i \ww_i + \ww_k \II_k \ww_k \Big) \Big|_\text{f}
 \nonumber \\
 & =  
 \FF_{ik}^{\text{f}} \cdot (\vv_i-\vv_k)
 + \ttt_{ik}^{\text{f}} \cdot \ww_{i} + \ttt_{ki}^{\text{f}} \cdot \ww_k 
 \nonumber \\
 & =  - f(u_{ik}^\perp, r_{ik}) \; u_{ik}^{\perp} \le 0,
 \end{align}
where $m_{i,k}$ are the masses, $\II_{i,k}$ the moments of inertia of
the two particles, and the subscript f refers to the fact that only
the effect of friction forces is considered.
No energy is dissipated in this model if $\uu_{ik}^\perp = 0$,
which corresponds to a rolling contact. Furthermore, the possibility
of spinning friction is also neglected. 

In addition to our analytical derivation, we have numerically verified
that energy is dissipated during particle collisions (Supplementary
Fig.~S1).

\subsubsection{Structure of corresponding stochastic force}
\label{subsec:fn}

Our goal is to construct stochastic forces such that, together
with the deterministic friction forces, the configurations 
are Boltzmann distributed at equilibrium. To this end, we examine
the corresponding Fokker-Planck equation.\cite{risken_book}

The system of interest consists of $N$ particles with positions
$\rr_i$, orientations characterized by two orthogonal unit vectors
$(\ee_i^{(1)},\ee_{i}^{(2)})$, momenta $\pp_i = m_i \vv_i$, and
intrinsic angular momenta $\jj_i = \II_i\ww_i$, which obey the
equations of motion $\dot{\pp}_i = \FF_i$, $\dot{\jj}_i = \ttt_i$,
$\dot{\rr}_i = \pp_i/m_i$, and $\dot{\ee}_i^{(\alpha)}= (\II_i^{-1}
\jj_i) \times \ee_i^{(\alpha)}$. Here $\FF_i$ and $\ttt_i$ denote the
total forces and torques acting on particle $i$.  The time evolution
of the $N$-particle distribution function ${\cal P}(\Gamma,t)$ with
$\Gamma = \{\rr_i,\pp_i,\ee_i^{(1,2)},\jj_i\}$ is described by a
Fokker-Planck equation $\partial_t {\cal P} = L_{\text{FP}} {\cal P}$.
We focus on the contributions of the deterministic tangential friction
force due to a frictional contact with particle $k$, 
$L_{ik}^{\text{f}}$, and the corresponding stochastic force,
$L_{ik}^{\text{R}}$, to the Fokker-Planck operator $L_{\text{FP}}$:
\begin{equation}
L_{\text{FP}} = \sum_{i < k}(L_{ik}^{\text{f}} + L_{ik}^{\text{R}})
+ \text{other contributions},
\end{equation}
where the sum $\sum_{i<k}$ runs over all pairs $(i,k)$ of particles 
with $i < k$.  The deterministic term accounts for the effect of the
friction force and torque, Eqs.\ (\ref{eq:transverse_force}) and
(\ref{eq:transverse_torque}), and can be decomposed into
\begin{equation}
L_{ik}^{\text{f}} = 
(\partial_{\pp_k} - \partial_{\pp_i}) \: {\bf J}_{ik,\pp}^{\text{f}}
- 
(\partial_{\jj_i} R_i+ \partial_{\jj_k} R_k) \: {\bf J}_{ik,\jj}^{\text{f}}
\end{equation}
with probability current operators 
\begin{align}
 {\bf J}_{ik,\pp}^{\text{f}} 
 =&
      \frac{f(u_{ik}^\perp,r_{ik})}{u_{ik}^\perp}
      \Big(
         \PP(\hr_{ik})\big(\frac{\pp_k}{m_k}-\frac{\pp_i}{m_i}\big) 
         \nonumber \\
         &\quad + \hr_{ik} \times 
           \big(\II_i^{-1} \jj_i R_i + \II_k^{-1} \jj_k R_k\big) 
            \Big),
\\
 {\bf J}_{ik,\jj}^{\text{f}} 
 =&
     \frac{f(u_{ik}^\perp,r_{ik})}{u_{ik}^\perp}
      \Big(\hr_{ik} \times 
       \big(\frac{\pp_k}{m_k}-\frac{\pp_i}{m_i}\big) 
       \nonumber \\
       &\quad - \PP(\hr_{ik}) 
         (\II_i^{-1} \jj_i R_i + \II_k^{-1} \jj_k R_k) 
     \Big).
\end{align}
We assume that it is possible to construct stochastic forces
and torques $\FF_{ik}^\text{R}$ and $\ttt_{ik}^\text{R}$ such that the
corresponding Fokker-Planck operator can be decomposed in the
same manner,
\begin{equation}
L_{ik}^{\text{R}} =
(\partial_{\pp_k} - \partial_{\pp_i}) \: {\bf J}_{ik,\pp}^{\text{R}}
- 
(\partial_{\jj_i} R_i+ \partial_{\jj_k} R_k) \: {\bf
J}_{ik,\jj}^{\text{R}}
\end{equation}
such that
\begin{eqnarray}
\label{eq:FP_current_pp}
 \Big(
 {\bf J}_{ik,\pp}^{\text{f}}  
 +{\bf J}_{ik,\pp}^{\text{R}}  
 \Big) \:
 {\cal P}_{\text{B}} &=& 0 \\
\label{eq:FP_current_jj}
 \Big({\bf J}_{ik,\jj}^{\text{f}}  \:
 +{\bf J}_{ik,\jj}^{\text{R}}  \:
 \Big) \:
 {\cal P}_{\text{B}} &=& 0.
\end{eqnarray}
for the Boltzmann distribution
\begin{equation}
\label{eq:p_stationary}
    {\cal P}_{\text{B}}(\Gamma) \propto
    \exp\Big[-\frac{1}{\kbT} \Big( U(\{\rr_i\}) 
    + \sum_i \Big(\frac{\pp_i^2}{2 m_i} 
    + \frac{1}{2} \jj_i \II_i^{-1} \jj_i \Big)\Big) \: \Big].
\end{equation}
In other words, we request that the Boltzmann distribution is not only 
the stationary solution of the equilibrium Fokker-Planck equation,
but also, that all probability currents 
${\bf J}_{ik,\pp}^{\text{R}} \: {\cal P}_{\text{B}}$ 
and ${\bf J}_{ik,\jj}^{\text{R}} \: {\cal P}_{\text{B}}$
vanish individually.

In general, the stochastic forces and torques will depend on the
instantaneous momenta and angular momenta of particles and generate
multiplicative noise. Hence the structure of the probability currents
depends on the choice of stochastic
calculus.\cite{paulbaschnagel_book} Here, we will discuss three
popular choices: (i) the It\^o calculus, which is commonly implemented
in numerical simulations, (ii) the Stratonovich calculus, which is
commonly used in physical theories because it preserves the chain
rule, (iii) the H\"anggi-Klimontovich calculus
\cite{haenggi1982stochastic,klimontovich1990ito,escudero2023ito}. Loosely
speaking, they differ in their definitions how to evaluate the
integrand in discrete implementations of time integrals, $\int \!\! f(t) \:
\ud t = \sum_i \big((1-\theta) f_i + \theta f_{i+1}\big) \big(t_{i+1}
- t_i\big)$.  The It\^o calculus specifies $\theta=0$, the Stratonovic
calculus $\theta = 1/2$, and the H\"anggi-Klimontovich calculus
$\theta=1$.  We can account for all three possibilities by making the
following Ansatz for the probability currents (see Section \ref{sec:A}):

\begin{align}
 {\bf J}_{ik,\pp}^{\text{R}} 
 =& \kbT \: D(u_{ik}^\perp,r_{ik})^{\theta}
      \Big(
         \PP(\hr_{ik})\big(\partial_{\pp_k}-\partial_{\pp_i}\big) 
         \nonumber \\
         & + \hr_{ik} \times 
           \big(\partial_{\jj_i} R_i + \partial_{\jj_k} R_k\big) 
            \Big) \: D(u_{ik}^\perp,r_{ik})^{1-\theta},
\label{eq:FP_prob_pp}
\\
 {\bf J}_{ik,\jj}^{\text{R}} 
 =& \kbT \: D(u_{ik}^\perp,r_{ik})^{\theta}
      \Big(\hr_{ik} \times 
       \big(\partial_{\pp_k}-\partial_{\pp_i}\big) 
       \nonumber \\
       & - \PP(\hr_{ik}) 
         (\partial_{\jj_i} R_i + \partial_{\jj_k} R_k) 
     \Big) \: D(u_{ik}^\perp,r_{ik})^{1-\theta}.
\label{eq:FP_prob_jj}
\end{align}
Using the general identities $\PP(\hr)^2 {\bm u} = \PP(\hr) {\bm u} =
- \hr \times (\hr \times {\bm u})$, and $\hr \times \PP(\hr) {\bm u} =
\PP(\hr)( \hr \times {\bm u}) = \hr \times {\bm u}$, which are valid
for all vectors ${\bm u}$, one can easily check that Eqs.\ 
(\ref{eq:FP_current_pp}) and (\ref{eq:FP_current_jj}) are fulfilled
with this choice, if the function $D(u,r)$ obeys the differential 
equation
\begin{equation}
\label{eq:ODE_D}
   \nu \: \kbT \: \partial_u D(u,r)
     - D(u,r) \: u = - f(u,r)
\end{equation}
with
\begin{equation}
 \label{eq:nu}
 \nu = (1- \theta) \: \left[ \big( \frac{1}{m_i} 
     +  \frac{1}{m_k} \big)
  +  \big(\frac{R_i^2}{I_i} +\frac{R_k^2}{I_k} \big) \right].
\end{equation}
The solution of Eq.\ (\ref{eq:ODE_D}) is
\begin{equation}
    \label{eq:dd_general}
        D(u,r) = \frac{1}{\kbT \nu} \: \int_u^\infty \ud u' \: 
          f(u',r) \: \ue^{-({u'}^2 - u^2)/2 \kbT \nu}
\end{equation}
for $\nu \neq 0$ (It\^o and Stratonovich case) and
\begin{equation}
    \label{eq:dd_linear}
        D(u,r) =f(u,r)/u
\end{equation}
for $\nu=0$ (H\"anggi-Klimontovich case).  We note that
(\ref{eq:dd_general}) reduces to (\ref{eq:dd_linear}) in the
special case of additive noise with $f(u,r) \propto u$.

The probability currents (\ref{eq:FP_prob_pp}) and (\ref{eq:FP_prob_jj})
correspond to the following random force and torque terms in the
stochastic differential equations of motion:
\begin{align}
    \label{eq:noise_force}
    \FF_{ik}^R &= - \FF_{ki}^R  =
       \sqrt{D(u_{ik}^\perp, r_{ik})} \: \Big[ 
         \PP(\hr_{ik})  \xx_{ik}^{\text{f}} 
            - \hr_{ik} \times \NN_{ik}^{\text{f}} \Big]\, , \\
            \label{eq:noise_torque}
    \frac{\ttt_{ik}^R}{R_i} &= \frac{\ttt_{ki}^R}{R_k} = 
    \sqrt{D(u_{ik}^\perp, r_{ik})} \:
       \Big[ \hr_{ik} \times \xx_{ik}^{\text{f}} + \PP(\ee_{ik})
       \NN_{ik}^{\text{f}} \Big]\ ,
\end{align}
where $\xx_{ik}^{\text{f}}$ and $\NN_{ik}^{\text{f}}$ are 
three-dimensional Gaussian white noise vectors with correlations
\begin{align}
    \langle \xx_{ij}^{\text{f}}(t) \xx_{kl}^{\text{f}}(t') \rangle
    =& \kbT \;
       \one \: (\delta_{ik} \delta_{jl} - \delta_{il} \delta_{jk}) \:
      \delta(t-t')
    \\
    \langle \NN_{ij}^{\text{f}}(t) \NN_{kl}^{\text{f}}(t') \rangle
    =& \kbT \;
       \one \: (\delta_{ik} \delta_{jl} + \delta_{il} \delta_{jk}) \:
      \delta(t-t').
\end{align}
A derivation of these expressions is provided in the Section \ref{sec:A}.

\subsubsection{Specific examples of friction types}

We will now discuss three particularly interesting choices
of amplitude functions $f(u,r)$. The first choice is 
the {\em linear friction} model:
\begin{equation}
    \label{eq:linear}
      f_{\text{l}}(u,r) = w(r) \: \gamma_{\text{f}} \: u,
\end{equation}
where the friction force depends linearly on the relative tangential
velocity with a distance-dependent factor $w(r)$ and a global
linear friction coefficient $\gamma_\text{f}$.
The integral (\ref{eq:dd_general}) then simply gives
\begin{equation}
        D_{\text{l}}(u,r) = w(r) \: \gamma_{\text{f}}
\end{equation}
and the noise is additive as discussed above.  In the linear case, the
combined deterministic friction and stochastic forces can be seen as
complicated dissipative particle dynamics (DPD)
thermostat\cite{Espanol1994Statistical}, which extends a transverse
DPD thermostat\cite{Junghans2008Transport} by terms that couple the
rotation and translation of particles. 
This thermostat has already been proposed by Espa\~nol in the 
context of fluid particle hydrodynamics \cite{espanol1998fluid}.

The second interesting choice is the {\em constant friction} 
or {\em Coulomb friction} model:\cite{elperin1997comparing}
\begin{equation}
    \label{eq:constant}
        f_{\text{c}}(u,r) = w(r)\: \kappa_{\text{f}}.
\end{equation}
Here the friction force only depends on the distance of
the particles with a prefactor $\kappa_\text{f}$. This
results in multiplicative noise, with the squared amplitude
having the form of a complementary error function, which can
    be evaluated very efficiently in numerical applications:
\begin{equation}
    D_{\text{c}}(u,r)= 
      \frac{w(r) \kappa_{\text{f}} \: \sqrt{\pi}}{\sqrt{2 \kbT \nu}} \:
    \ue^{ \frac{ u^2}{2 \kbT \nu}} \: 
    \text{Erfc}\big( \frac{u}{\sqrt{2 \kbT \nu}} \big)
\end{equation} 
This model works well in practice as we shall show below,
but it has the conceptual problem that the friction force
does not have a well-defined limit at $u \to 0^{\pm}$, 
and may rapidly oscillate between finite positive and negative
forces for small tangential velocities.
 
The third example, the {\em Coulomb-Newton friction} model
avoids this problem and is popular in granular matter 
models \cite{elperin1997comparing}:
\begin{equation}
    \label{eq:cn}
      f_{\text{CN}}(u,r)=
        \min\big[\gamma_{\text{f}} \: u,w(r) \: \kappa_{\text{f}}\big]
\end{equation}
In this case, the function $D_{\text{CN}}(u,r)$ is given by
    \begin{align}
        \lefteqn{D_{\text{CN}}(u,r) } \nonumber \\
        &= \left\{
        \begin{array}{ll} 
          \frac{w(r) \kappa_{\text{f}} \: \sqrt{\pi}}{\sqrt{2 \kbT \nu}} \:
           \ue^{ \frac{ u^2}{2 \kbT \nu}} \: 
          \text{Erfc}\big( \frac{u}{\sqrt{2 \kbT \nu}} \big)
           &: u \ge \frac{w(r)\kappa_{\text{f}}}{\gamma_\text{f}} \:  
           \\ 
           \gamma_\text{f}\:
           \Big( \: 1 - 
             \ue^{\big(u^2-(\frac{w(r)\kappa_{\text{f}}}{\gamma_\text{f}})^2
                         \big)/2\kbT \nu} \Big)
           \\ \quad  
            + 
          \frac{w(r) \kappa_{\text{f}} \: \sqrt{\pi}}{\sqrt{2 \kbT \nu}} \:
           \ue^{ \frac{ u^2}{2 \kbT \nu}} \: 
          \text{Erfc}\big( 
            \frac{w(r) \kappa_{\text{f}}
                /\gamma_{\text{f}}}{\sqrt{2 \kbT \nu}} \big)
            &: u < \frac{w(r)\kappa_{\text{f}}}{\gamma_\text{f}}
        \end{array} 
        \right.
    \end{align}
In all three examples, $w(r)$ is a smooth function of the
center-to-center distance between the particles $i$, and $k$, which
ensures that the friction vanishes if the particles are far apart.  In
the classical Coulomb friction model, the tangential friction force is
assumed to be  proportional to the normal (repulsive) conservative
force between the particles. In the applications shown below, we will
consider particles that interact via a purely repulsive
Weeks-Chandler-Anderson (WCA) potential\cite{WCA}, 
\begin{align}
     U_\text{WCA}(r)=
      \begin{cases}
     4 \epsilon \left[ \left( \frac{\sigma}{r} \right)^{12} 
       - \left(\frac{\sigma}{r} \right)^{6} \right] 
       + \epsilon &,\, r \leq \sqrt[6]{2}  \sigma \\
     0 & ,\, r > \sqrt[6]{2}\sigma \,.
     \end{cases}
     \label{eq:wca}
 \end{align}
Therefore, adopting the Coulomb friction Ansatz, we will choose
$w(r) = - \ud U_\text{WCA}(r)/\ud r$. 

To validate our approach, we have performed three-dimensional
simulations of systems of identical passive colloids interacting with
repulsive WCA interactions and frictional contact interactions of the
Coulomb-Newton type (with $R=\sigma/2$ and $\kbT = \epsilon$).  No
additional thermostat is applied.  We use a simple Euler-forward
scheme to integrate over stochastic forces, corresponding to an
implementation of It\^o calculus.  Fig.~S2 shows the resulting
distributions of the speeds and angular speeds of particles. They are
in perfect agreement with the Maxwell-Boltzmann distribution. 

\subsection{Application examples}
\label{sec:applications}

In the following, we investigate the effect of frictional contact
forces in three different systems: First, in
Section~\ref{sec:thermal}, we simulate simple passive fluids to
demonstrate the thermodynamic inconsistencies that arise if only the
deterministic frictional contact forces are included in a model.
Second, in Section~\ref{sec:poiseuille}, we consider pressure-driven
Poiseuille flows in a slit channel and investigate the coupling of
velocity and angular velocity gradients due to frictional contacts in
such channels.  Furthermore, we use the flow profiles to determine the
dynamic viscosity arising from frictional contact interactions.  Last,
in Section~\ref{sec:mips}, we study the influence of frictional
contacts on motility-induced phase separation (MIPS) in systems of
active particles.  

All particles have identical masses $m$, radii $R =\sigma/2$,
and moments of inertia $\II = \frac{2}{5} m R^2 \mathbf{1}$, 
interact with WCA interactions (Eq.\ (\ref{eq:wca})), and may be
coupled to a medium at rest. The basic equations
of motion read
\begin{align}
    \label{eq:vv}
     m \dot{\vv}_i =&  \FF_i^{\text{ne}}
     - \gamma \vv_i + \xx_i^{\text{s}} \nonumber \\
                     & +  \sum_{k\neq i} (-\nabla_i U_\text{WCA}(r_{ik}) 
                     + \FF_{ik}^{\text{f}} + \FF_{ik}^\text{R}) \\
    \label{eq:ww}
    \frac{\ud}{\ud t} (\II \: \ww_i) 
      =& 
     \ttt_i^{\text{ne}}  
      - \gamma_R \ww_i + \NN_i^{\text{s}}
      + \sum_{k\neq i}(\ttt_{ik}^{\text{f}} + \ttt_{ik}^\text{R})\, \\
    \label{eq:ee}
    \dot{\ee}_i^{(\alpha)} =& \ww_i \times \ee_i^{(\alpha)}
     \quad \text{for $\alpha = 1,2$}\\
    \label{eq:rr}
    \dot{\rr}_i =& \vv_i
\end{align}
where $\gamma$ and $\gamma_R$ are translational and rotational
friction coefficients describing the interaction with the
medium, and the forces $\xx_i^{\text{s}}$ and
torques $\NN_i^{\text{s}}$ are uncorrelated Gaussian white noise terms
with zero mean and variance
\begin{align}
    \langle \xx_i^{\text{s}}(t) \xx_k^{\text{s}}(t') \rangle 
       &= 2 \kbT \: \gamma \: \one \: \delta_{ik} \:\delta(t-t' ) \\
    \langle \NN_i^{\text{s}}(t) \NN_k^{\text{s}}(t') \rangle 
       &= 2 \kbT \: \gamma_\text{R} \: \one \: \delta_{ik} \: 
          \delta(t-t' ),
\end{align}

The forces $\FF_{ik}^{\text{f/R}}$ and torques
$\ttt_{ik}^{\text{f/R}}$ model interparticle frictional contacts and
as described in the previous section.  Additional forces
$\FF_i^{\text{ne}}$ and torques $\ttt_i^{\text{ne}}$ may drive the
system out of equilibrium.  All densities are expressed by the area
fraction $\rho_\text{area}=\pi N R^2/A$, the volume fraction
$\rho_\text{volume}=4\pi N R^3/3V$, or the number density $n=N/A$,
where $N$ is the number of particles, and $A$, $V$ are the area and
the volume. In two-dimensional systems, the position of the particles
is restricted to a plane. 

All quantities will be given in units of the length $\sigma$, the
thermal energy $\kbT$, and a ``natural'' time unit $\tunit$, which
is specified for each case study, along with other simulation
parameters, in the Methods section.

\label{sec:model}

\FloatBarrier

\subsubsection{Thermodynamic consistency in models 
with frictional contact interactions}
\label{sec:thermal} 

In Section \ref{sec:theory}, we have shown that dissipative frictional
forces necessarily reduce the kinetic energy of colliding particles,
but combined dissipative and stochastic frictional forces constitute a
valid thermostat in simulations of passive colloidal systems (see
Supplementary Fig.~S2). When considering systems of particles in an
implicit medium as in Eqs.\ (\ref{eq:vv}) and
(\ref{eq:ww}), one might be tempted to omit the stochastic force and
torque terms, $\FF_{ik}^\text{R}$ and $\ttt_{ik}^\text{R}$, assuming
that the coupling to a bath medium (i.e., the
corresponding Langevin thermostat) is sufficient to guarantee sampling
of a canonical ensemble in the absence of nonequilibrium forces
$\FF_i^{\text{ne}}$.  In the following, we will demonstrate that this
assumption is not correct, and that the stochastic frictional contact
terms must necessarily be included to avoid thermodynamic
inconsistencies.

For this purpose, we have investigated two and three-dimensional
systems of passive particles with frictional contact interactions,
which are coupled to a bath medium as described above. The
nonequilibrium forces and torques were set to zero,
$\FF_i^{\text{ne}} =  \ttt_i^{\text{ne}} = 0$.  The area fraction of
the two-dimensional system was chosen $\rho_\mathrm{area}=0.6$ and the
volume fraction of the three-dimensional system
$\rho_\mathrm{volume}=0.3$. All systems contain $N=10\,000$ particles
with periodic boundary conditions in all directions.  We performed
simulations for the friction types linear, Coulomb and Coulomb-Newton.
Each type was simulated twice, once without stochastic friction
$\FF_{ik}^\text{R}$ and torques $\ttt_{ik}^\text{R}$ and once
including them. The systems were equilibrated over a time $20\,\tunit$
using a time step of $\Delta t = 0.001\,\tunit$ for the linear and
Coulomb-Newton friction model and $\Delta t = 0.0001\,\tunit$ for the
Coulomb friction model. The data were then collected over a simulation
time period of $200\,\tunit$.  The distributions of speeds and angular
speeds were then compared to the theoretically expected
Maxwell-Boltzmann distributions at the bath temperature $\kbT$. 

\begin{figure}
    \centering
    \includegraphics[width=0.4\textwidth]{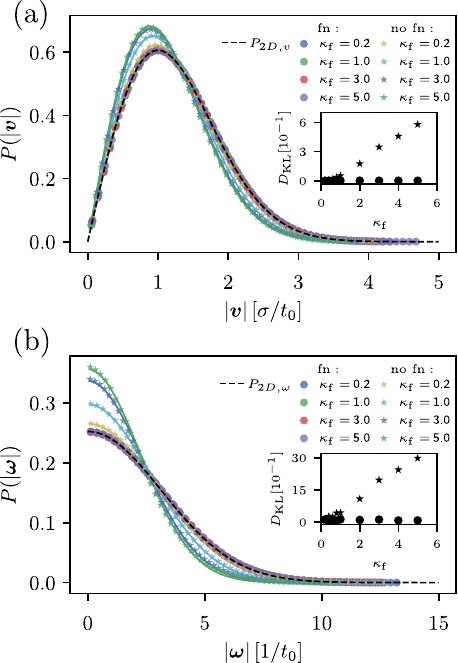}
    \caption{(a) Speed histograms and (b) angular speed histograms for
    a two-dimensional passive colloid system with frictional
    contact interactions according to the Coulomb friction model, Eq.\
    (\protect\ref{eq:constant}), for different friction constants
    $\kappa_f$ as indicated.  Stars (labelled "no fn") show data
    from simulations without stochastic contact friction terms,
    circles (labelled "fn") data for the full model. Solid
    colored lines are fits to appropriate Maxwell-Boltzmann
    distribution with fitted effective temperature, $T_\text{eff}$,
    black solid line gives the theoretical expectation for
    $\kbT_\text{eff} = \kbT$.  The inset shows the corresponding
    Kullback-Leibler divergence between the distribution measured in
    simulations and the fitted Maxwell-Boltzmann distribution.
    \label{fig:Histo_const_2D}}
\end{figure}

\begin{figure}
    \centering
    \includegraphics[width=0.5\textwidth]{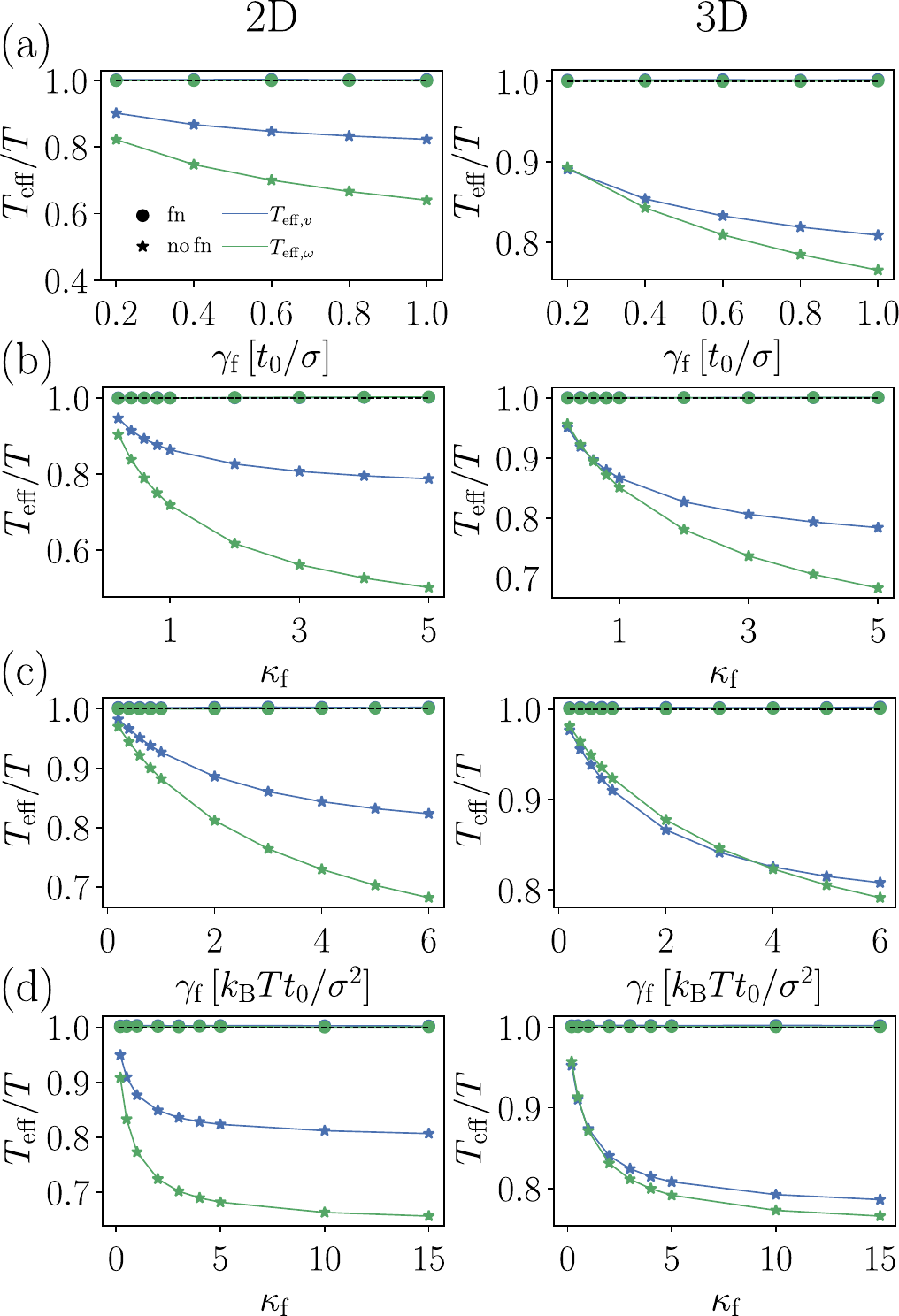}
    \caption{$T_\mathrm{eff}$ Effective kinetic temperatures obtained
    by fitting speed (green) and angular speed (blue) histograms to
    the corresponding Maxwell-Boltzmann distributions from simulations
    without (stars) and with (circles) stochastic contact friction
    terms in two-dimensional (left) and three dimensional (right)
    systems and for different friction types: (a) linear friction
    model (b) Coulomb friction model, (c,d) Coulomb-Newton friction
    model with (c) fixed $\kappa_\text{f}=5$ and varying
    $\gamma_\text{f}$, and (d) fixed $\gamma_\text{f}=6 \: \kbT
    \tunit/\sigma^2$ and varying $\kappa_\text{f}$. \label{fig:T_eff}}
\end{figure}

Fig.~\ref{fig:Histo_const_2D} depicts such a comparison exemplarily
for a two-dimensional system with frictional contacts according to the
Coulomb friction model and different values of $\kappa_\mathrm{f}$.
Corresponding figures for other types of friction and in three
dimensions are provided in supplemental information (Supplementary
Fig.~S3 and Fig.~S4). When omitting the stochastic contact forces and
torques in the simulations (star symbols), the speed and angular speed
distributions clearly deviate from the theoretically expected
Maxwell-Boltzmann distribution $P_\mathrm{2d,v}$ and
$P_\mathrm{2d,\omega}$, which is shown as dashed black line. The
distribution is shifted to the left, such that smaller speeds are
overrepresented and larger ones are underrepresented, indicating that
the Langevin thermostat is not able to fully compensate the loss of
kinetic energy due to frictional forces during collisions. However,
when the stochastic contact terms are included, the simulation data
(circles) coincide with the expectation.

One might argue that the main effect of omitting the stochastic
contact terms is to reduce the kinetic temperature of the particles,
but that the system might still adopt an ``equilibrium'' state
corresponding to such a reduced temperature. However, a closer look at
the data shows that this is not the case. First, we can determine the
``effective'' temperature $T_\text{eff}$ by fitting the speed and
angular speed histograms to the Maxwell-Boltzmann distribution, using
$T_\text{eff}$ as fit parameter. The result is shown in
Fig.~\ref{fig:T_eff} for two- and three-dimensional systems, three
different types of contact friction, and varying friction parameters.
Not surprisingly, the effective temperature decreases with increasing
friction parameters.  More importantly, however, the effective kinetic
temperature obtained from the speed histogram and the angular speed
histogram differ from each other. This clearly indicates that the
system cannot be mapped to an equilibrium system.

A second, more subtle nonequilibrium signature is that the shape of
the distribution also differs from the expected Maxwell-Boltzmann
shape. To quantify this, we have calculated the Kullback-Leibler
divergence, $D_\text{KL}$, between the real (simulated) histograms and
the fitted histograms with shifted effective temperature: 
\begin{equation}
D_\text{KL} \big(P_\text{sim}(\bm{x})||P_\text{fit}(\bm{x})\big) =
\int \!\!\! \ud \bm{x} \: P_\text{sim}(\bm{x}) \: \ln\Big(
\frac{P_\text{sim}(\bm{x})}{P_\text{fit}(\bm{x}}\Big),
\end{equation}
where $\bm{x}$ stands for $\vv$ or $\ww$, $P_\text{sim}(\bm{x})$ is
the distribution from simulation, and $P_\text{fit}(\bm{x})$ the
fitted distribution. The results for the Coulomb friction model are
shown in the insets of Fig.~\ref{fig:Histo_const_2D}. If the
stochastic contact terms are omitted in the simulation, the
Kullback-Leibler divergence deviates from zero for large friction
parameters, indicating that the shape of the distribution from the
simulations deviates from the fitted Maxwell-Boltzmann distribution.
If the stochastic terms are included, the Kullback-Leibler divergence
is zero, hence the simulations produce the correct distribution
functions.

\FloatBarrier

\subsubsection{Pressure-driven flows}

\label{sec:poiseuille} 

Next, we study the rheological properties of fluids of repulsive
particles with frictional surface interactions. Since individual
particles are not coupled to a substrate, the parameters $\gamma$ and
$\gamma_R$ in Eqs.  (\ref{eq:vv}, \ref{eq:ww}) are set to zero.
\rev{Also, solvent-mediated hydrodynamic interactions between
particles are neglected\cite{brady1988stokes}.} 
However, the particles may still interact with confining walls.
\rev{This model describes, e.g., colloidal suspensions in a regime
where the total shear viscosity is dominated by particle-particle
interactions rather than particle-solvent friction.}
We consider a Poiseuille flow in a slit channel geometry.
The nonequilibrium forces are set to $\FF_i^\text{ne} \equiv F \:
\ee_x$, i.e., all particles are subject to a constant bulk force in
the $x$ direction.  Such a setup is commonly used to mimic
pressure-driven flows\cite{Keaveny2005} in a pressure gradient $\ud
P/\ud x = - F \: n$, where $n$ is the local number density of
particles.  The particles are confined by two identical walls,
consisting of a layer of virtually immobilized particles  $j$ (mass
$10^{10}\,m$) with diameter $\sigma$ and $z$-coordinates
\begin{equation}
y_j = \pm 8.5847\sigma + \delta_w \: \zeta_j,
\label{eq:wall}
\end{equation}
which are arranged on a square lattice with lattice constant
$a=\sigma$ in the $(xz)$-plane. In Eq.\ \ref{eq:wall}, $\zeta_j$
denotes uncorrelated random numbers that are drawn from a Gaussian
distribution with mean zero and variance 1, and the parameter
$\delta_w>0$ allows to tune the degree of surface roughness.  For
technical reasons, the random numbers are the same on both sides of
the slab. All particles, mobile and immobile, interact via repulsive
WCA interactions and frictional contact interactions of the
Coulomb-Newton type (Eq. \ref{eq:cn}). We considered systems of 10000
particles at volume fraction $\rho_\text{volume}=0.15$ in a
simulation box with accessible volume $V= (151.694 \times 15.1694
\times 15.1694) \sigma^3$, and periodic boundary conditions in the $x$
and $z$ directions.  The time step in the simulations was $\Delta t =
0.001\,\tunit$.  After initialization, each system was simulated over
a time of $500\,\tunit$, which was sufficient to reach a stationary
state (Supplementary Fig.~S5). Data were then collected over a
simulation time period of $6000\,\tunit$. In the following, we will
show data obtained for bulk forces $F=0.01 \kbT/\sigma$. We have also
simulated systems at $F=0.02 \kbT/\sigma$ and verified that the
velocity and angular velocity profiles are proportional to $F$ as
expected in the linear regime.

\begin{figure}
    \centering
    \includegraphics{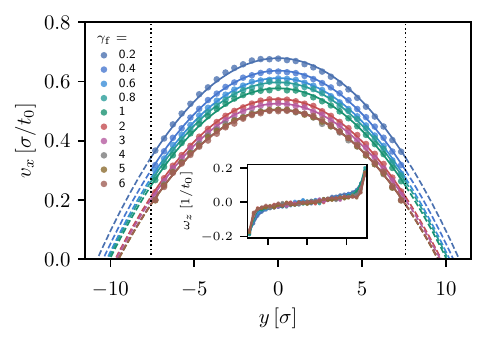}
    \caption{Examples of velocity profiles for fluids with frictional
contact interactions of the Coulomb-Newton type with parameter
$\kappa_\text{f}=5$ and different values of $\gamma_\text{f}$ as
indicated, which are confined between walls with roughness parameter 
$\delta_w=0$ (see text) and subject to a bulk force $F=0.01 \:
\kbT/\sigma$. Solid lines show fit to a parabolic profile. Inset 
shows corresponding profiles of the $z$-component of the angular velocity. 
}
\label{fig:poiseuille}
\end{figure}

Fig.\ \ref{fig:poiseuille} (main figure) shows examples of resulting
velocity profiles for different thermostat parameters
$\gamma_\text{f}$ at roughness parameter $\delta_w=0$.  They have the
parabolic shape expected for Poiseuille flow of Newtonian fluids, if
one allows for substantial slip at the surface.  Compared to regular
DPD simulations of Poiseuille flow, in our system, slip is facilitated
by the possibility that particles roll off on the surface without
dissipating energy due to friction.  This rolling involves a
collective rotational motion to the particles, which however quickly
decreases further away from the surface.  More generally, the inset
of Fig.~\ref{fig:poiseuille} (main figure) demonstrates that shear in
the velocity profiles $(\ud v_x/\ud v_y \neq 0)$ induces rotational 
motion of particles in the fluid.

Fitting the velocity profiles in Fig.~\ref{fig:poiseuille} to the
prediction of the Stokes equation for a fluid with volume fraction
$\rho_\text{volume}$, and partial slip boundary conditions
($v_x(y_\text{B})/v'_x(y_\text{B})= \delta_\text{B}$, 
where $\delta_\text{B}$ is the slip length, and $z_\text{B}$ 
the position of the hydrodynamic boundary)
\begin{equation} 
\label{eq:poiseuille}
v_x(y) = - \frac{3\rho_{\text{volume}}}{\pi\eta} F (y^2 - y_{_0}^2),
\end{equation} 
we can extract the dynamic viscosity $\eta$ and the zero crossing
parameter $y_{_0}=(\delta_\text{B}+ y_\text{B})^2 -
\delta_\text{B}^2$. The parameters $\delta$ and $y_\text{B}$ can be
disentangled following a procedure proposed by
Allen\cite{smiatek2008tunable}, which is based on the comparison of
Poiseuille flow profiles and complementary Couette flow profiles. To
apply it, we performed additional simulations with a slightly altered
setup. We removed the constant bulk force and slid the walls in
opposite $x-$directions at a constant velocity
$v_\text{wall}=\pm1\,\sigma/\tunit$, which yielded linear velocity
profiles (see Supplementary Fig.~S6). Then we determined the distance
$C$ between the two points where the extrapolated linear profile
equals the wall velocity. From this we can determine the squared slip
length via $\delta^2_\text{B}=C^2/4-y{_0}^2$ and the distance
between the physical and the hydrodynamic boundary as
$y_\text{B}=\delta_\text{B}-(C-15.1694)/2$.

\begin{figure}
    \centering
    \includegraphics[width=\linewidth]{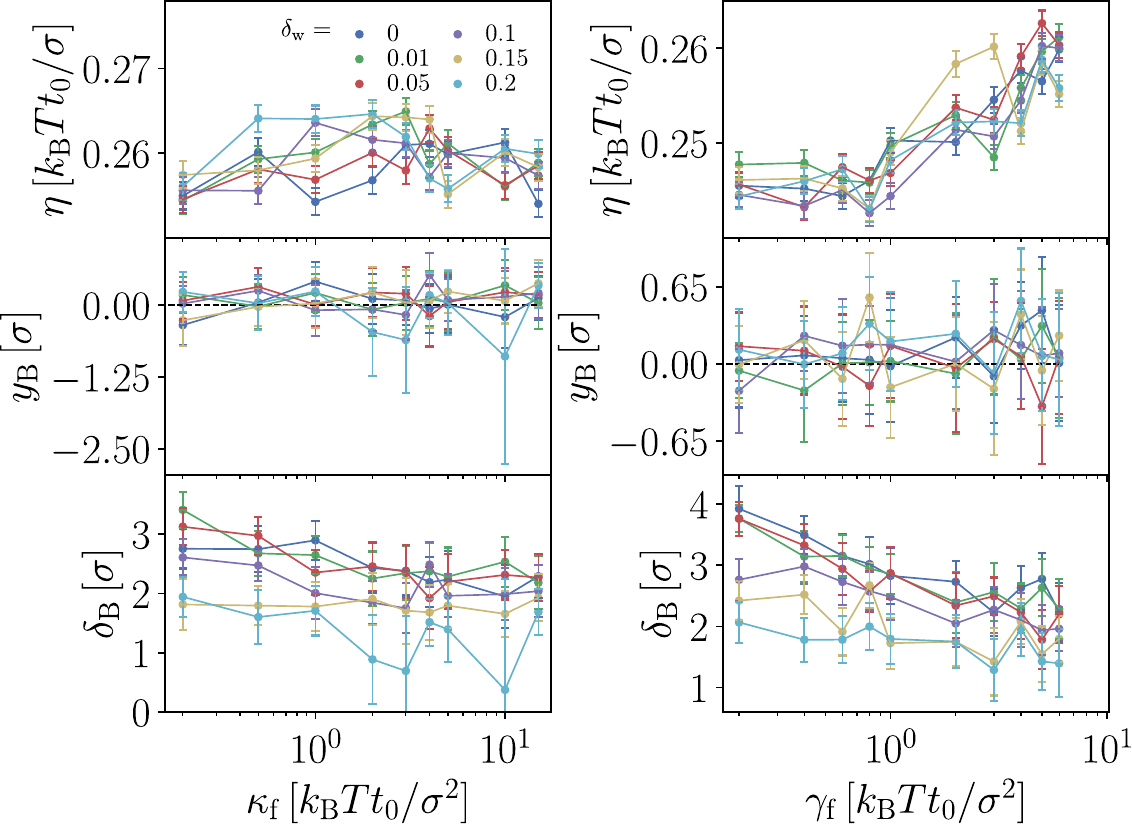}
    \caption{ Dynamic viscosity $\eta$ (top), distance between the
    hydrodynamic boundary and the physical boundary $y_\text{B}$
    (middle), and slip length $\delta_\text{B}$ (bottom) as obtained
    from parametric fits to Poiseuille flow and Couette flow
    simulation profiles as in Fig.\ \protect\ref{fig:poiseuille} and
    Fig.\ \protect S6 of fluids with frictional contact interactions
    of the Coulomb-Newton type in slit geometry. The results are shown
    for various values of the surface roughness parameter $\delta_w$
    as functions of the friction parameter $\kappa_\text{f}$ with
    fixed $\gamma_\text{f} = 6\: \kbT \tunit/\sigma^2$  (left), and
    $\gamma_\text{f}$ with fixed $\kappa_\text{f} = 5$ (right). Error
    bars represent the one standard deviation uncertainties of the
    fitted parameters.}
    \label{fig:viscosity_slip}
\end{figure}

The results of such fits are shown in Fig.~\ref{fig:viscosity_slip}.
The distance between the physical and the hydrodynamic boundary
$y_\text{B}$ equals zero (within the error), compatible with findings
from previous studies of simple fluids at hard walls
\cite{smiatek2008tunable,zhou2012anisotropic}.  Interestingly, the
dynamic shear viscosity of the fluid only slightly increases with the
friction parameter $\gamma_\text{f}$ and is independent of the
parameter $\kappa_\text{f}$, indicating that, at this volume fraction,
the shear viscosity is basically determined by the conservative
interactions between the particles. In contrast, the contact friction
does influence the surface slip: The stronger the contact friction
interactions, the smaller the slip length. Nevertheless, it seems
difficult to achieve zero slip length in this system due to the
possibility that particles can roll off at the surface. The slip
length is roughly independent of the roughness parameter $\delta_w$ up
to $\delta_w \sim 0.1$ and then decreases as $\delta_w$ is increased
further

For comparison, we have also studied the effect of turning off 
the frictional contact noise. The results are shown in
Supplementary Fig.~S7. In the absence of thermal noise, the
viscosity drops roughly by a factor of 5, and likewise, the surface slip  
increases dramatically. Hence thermal noise significantly 
enhances the energy dissipation in the fluid. The reason is that 
particles in noiseless fluids almost stop moving perpendicular to the 
bulk force or sliding walls, resulting in fewer particle collisions 
and fewer frictional contacts.
 
Even though our results so far suggest that the flow profiles can be
described by the Stokes equation, the relatively strong coupling to
angular rotation motion suggests that nonlinear rheological effects
might be important in fluids involving frictional contact
interactions. Unfortunately, increasing the driving force $F$ and
pushing it into the nonlinear regime mainly has the effect of
increasing the slip length, such that the flow profiles become plug
like (data not shown). In order to avoid this effect and investigate
the coupling between shear rate and rotational motion of particles at
higher shear rates, we have used an artificial simulation setup which
is popular in rheological simulation studies of complex
fluids\cite{backer2005poiseuille,fedosov2010steady}: We apply periodic
boundary conditions in {\it all} directions, but split the simulation
box in half in the $y$ direction and impose opposing bulk forces in
both halves: $\FF_i^\text{ne} = F \: \ee_x$ for $y_i >0$, and
$\FF_i^\text{ne} = - F \: \ee_x$ for $y_i <0$.  According to the
Stokes equation, this should result in a sequence of two parabolic
flow profiles in opposite directions. In our simulations, we found
that, instead, vortices developed in the high shear regions at higher
forces bulk forces, which eventually destroyed the stationary
Poiseuille-like flow profiles. An example for $F=0.5 \: \kbT/\sigma$
is provided in Supplementary Movie 1. The formation of these vortices
was prevented in the presence of solid walls. Nevertheless, our
results suggest that the linear response regime in fluids with
frictional contact interactions is small and that their dynamic
behavior might often be dominated by nonlinear effects.

\FloatBarrier

\subsubsection{Active particles and motility-induced phase separation}
\label{sec:mips}

As a final example, we consider the effect of frictional contact
interactions on structure formation in systems of active particles.
Specifically, we focus on the phenomenon of motility-induced phase
separation (MIPS) \cite{Cates_2015}, which has been observed both in
systems of active Brownian particles (ABPs) and active Langevin
particles (ALPs). ALPs follow the equations of motion
(Eqs.~\ref{eq:vv}-\ref{eq:rr}) with effective active force
$\FF_i^\text{ne} = F_0 \ee_i^{(1)}$ (and originally without frictional
contact terms)\cite{ABP_ALP}. ABPs obey the same equations of motion
as ALPs in the overdamped limit (Eqs.~\ref{eq:vv}-\ref{eq:rr} with $m
\to 0$ and $\II \to 0$).  In a certain parameter range, the effective
active force can induce phase separation between coexisting dense and
dilute regimes even in the absence of attractive conservative forces
between particles. The reason is, loosely speaking, that particles
which happen to push against each other by the active forces remain
stuck to each other for some time before they reorient. If other
particles collide with the stuck particle assembly before the
particles reorient, the clusters tend to grow until the system is
phase separated. Roughly, this happens at sufficiently high active
particle speeds and densities, such that the collision rate is higher
than the reorientation rate.  Frictional contact forces can enhance
particle aggregation, as two particles with zero or small initial
angular velocity that collide off-center (nonzero impact parameter)
are forced to rotate in the same direction, causing them to stick
together even longer.  This effect has recently been reported for
systems of ABPs by Nie et al.\cite{nie2020frictional} in a model that
accounted for dissipative friction contact forces similar to the
Coulomb-Newton type and which is also well-known in the granular
community \cite{herrmann1998modeling}. The model by Nie et al. did not
include corresponding stochastic terms with thermodynamically
consistent coupling of translational and rotational noise.

\begin{figure}
    \centering
    \includegraphics[width=.5\linewidth]{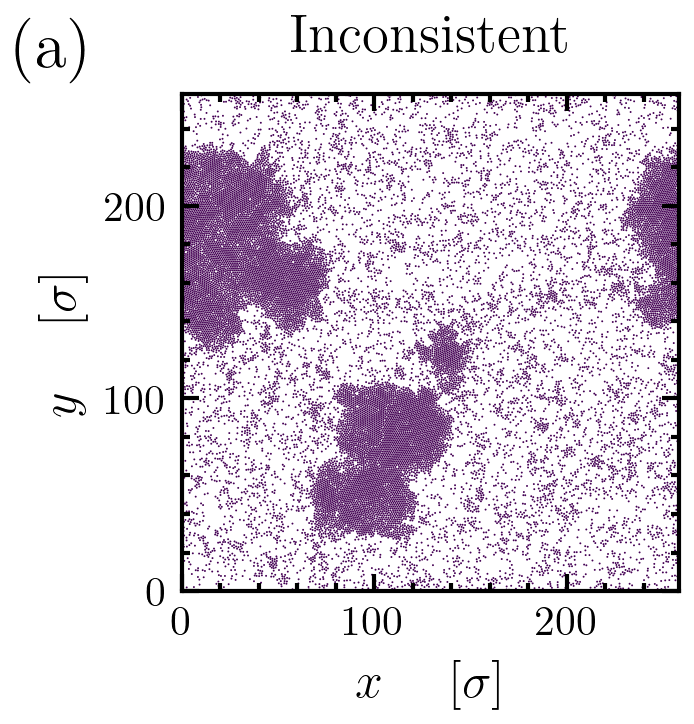}%
    \includegraphics[width=.5\linewidth]{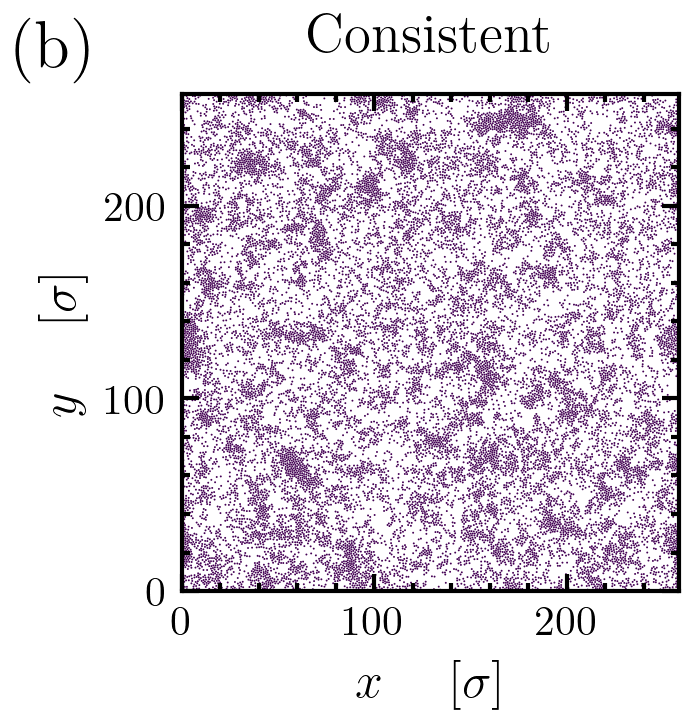}
    \caption{Snapshots of ALP simulations of 20000 Particles after a
    sufficient equilibration period with Coulomb-Newton friction. (a)
    Thermodynamically inconsistent treatment and (b) thermodynamically
    consistent treatment of the frictional contacts.  In the
    thermodynamically consistent treatment of frictional contacts, no
    MIPS occurs, whereas on the left MIPS can be seen.  The parameters
    of the simulations are marked with a red circle in
    Fig.~\ref{fig:mips}.  }
    \label{fig:mips_snapshot}
\end{figure}

\begin{figure}
    \centering
     \includegraphics[width=1\linewidth]{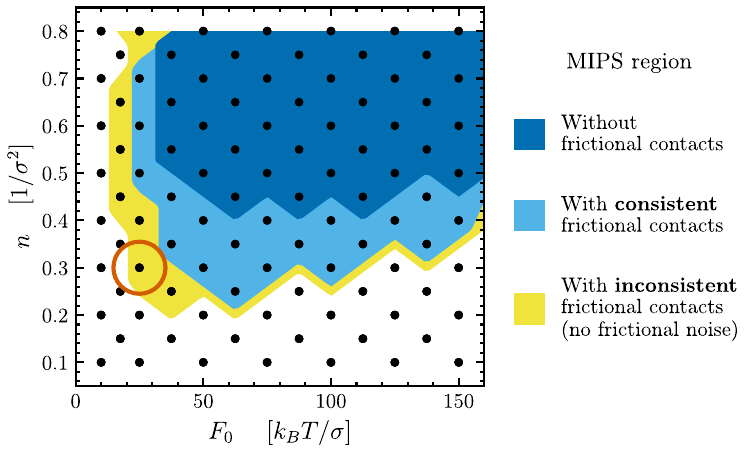}
    \includegraphics[width=1\linewidth]{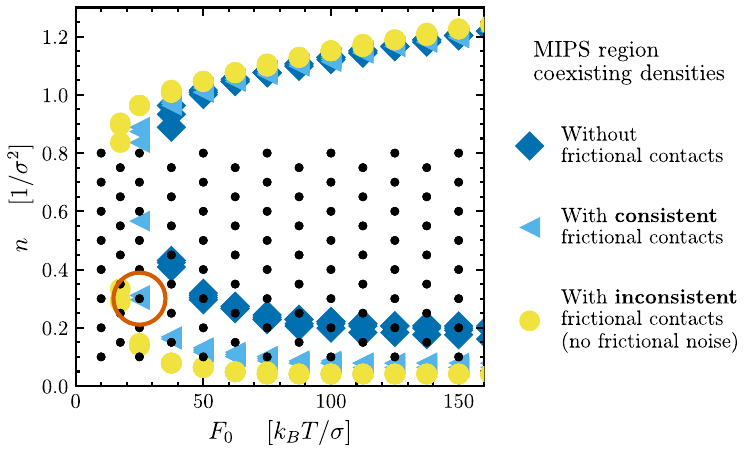}
    \caption{ Top: MIPS regime in two-dimensional systems of ALPs
    without frictional contacts (dark blue), with thermodynamically
    incorrect frictional contacts (yellow) as well as correct
    frictional contacts (light blue) depending on active force $F_0$
    and number density $n$.  The black dots show where we have
    performed simulations and the colored regions show where MIPS
    spontaneously emerges (bimodal density distribution) from a
    uniform initial state in our simulations.  The red circle marks
    the simulation parameters of the simulation snapshots in
    Fig.~\ref{fig:mips_snapshot}.  Bottom: Corresponding coexisting
    densities in the MIPS regime.  See text for further explanation
    and simulation details.  }
    \label{fig:mips}
\end{figure}

In our simulations, we study two-dimensional systems of $N=20,000$
ALPs with model parameters as specified earlier (beginning of Section
\ref{sec:model}), which were chosen in accordance with own previous
work on MIPS where we know the parameter regime of the MIPS
region\cite{Hecht2022}. In particular, the mass is set to be $m = 0.05
\:\kbT \: (t_\text{R}/\sigma)^2$, which is still in the underdamped
regime as compared to the overdamped system studied by Nie et al.
\cite{nie2020frictional}. We include frictional contact interactions
of the Coulomb-Newton type with parameters $\gamma_\text{f} = 1 \cdot
\kbT t_\text{R}/\sigma^2$ and $\kappa_\text{f} = 1$.  The simulation
time step was $\Delta t = 10^{-5} \cdot t_\text{R}$, and periodic
boundary conditions were applied in all directions.  Neglecting the
thermodynamic correction of the frictional contacts can cause the
collective behavior to vary significantly. This can be seen in
Fig.~\ref{fig:mips_snapshot}, where the simulation with
thermodynamically inconsistent frictional contacts leads to MIPS,
whereas the simulation of the consistent treatment shows no MIPS.  To
explore the importance of a thermodynamically consistent treatment
more systematically, we now compare the phase diagram for three
different cases: (i) without frictional contacts, (ii) with
Coulomb-Newton friction with frictional noise, and (iii) with
Coulomb-Newton friction, but without frictional noise. MIPS was
identified using AMEP\cite{hecht2025amep} by computing the local
density on a uniform grid.  A bimodal distribution in the resulting
density histogram indicates phase separation.  The coexisting
densities within the MIPS region \cite{Bialke2015, Levis2017} were
identified by locating the two peaks in this density histogram (cf.
Fig.~\ref{fig:mips}, bottom).

The resulting non-equilibrium MIPS phase diagrams for systems with and
without frictional interactions are shown in Fig.~\ref{fig:mips}
(top).  Interestingly, the coexistence densities -- which in an
equilibrium system would give the binodals of phase separation -- do
not coincide with the boundaries of the MIPS regime. This discrepancy
might be a signature of the system's nonequilibrium nature or may
arise from strong finite-size effects in the simulations.

As reported by Nie et al.\cite{nie2020frictional} for ABPs, the
contact friction forces significantly expand the regime in which MIPS
is observed.  This can also be seen in the case of the inertial ALPs
we investigated.  More importantly, the comparison between the
thermodynamically consistent case (with frictional noise) and the
thermodynamically inconsistent case (no frictional noise) shows a
notable difference in the morphology of the MIPS phase diagram. This
demonstrates that neglecting frictional noise, despite altering the
equilibrium distribution of colloidal particles only relatively
mildly, can lead to a wrong prediction for the collective behavior of
a nonequilibrium many particle system in certain parameter regimes.



\section{Conclusion}

In sum, we have  constructed a thermodynamically consistent way of
modeling thermal fluctuations in particle models with frictional
contact interactions. Our approach is based on the construction of
appropriate Fokker-Planck equations for such systems under the
constraint that the stationary solution for equilibrium systems should
be the Boltzmann distribution function. Importantly, it shows that
coupled frictional forces and torques arising from frictional contacts
must necessarily be accompanied by stochastic forces and torques that
are coupled in a structurally similar manner. We have demonstrated
that neglecting such stochastic terms in equilibrium simulations
results in stationary distributions that differ from the
Boltzmann distribution, both regarding effective temperature and
shapes of the distribution. Even more seriously, effective
temperatures related to different degrees of freedom, such as
momenta and angular momenta, are no longer equal if the
stochastic forces and torques are not modelled consistently.

We have considered a general form of frictional contact interactions,
which is inspired by models for granular matter. In general, the
resulting stochastic terms depend on the velocities and angular
velocities of particles and are therefore instances of multiplicative
noise. Hence, we have separately considered three popular choices of
stochastic calculus, It\^o, Stratonovich, and H\"anggi-Klimontovich,
and we have provided explicit expressions for the stochastic Langevin
noise terms for all three choices. Then we have validated the approach
with simulations based on the It\^o calculus and presented
applications to driven flows and active particles undergoing MIPS.

In driven flows, we have seen that frictional contacts lead to a
coupling of rotational and translational degrees of freedom, which may
lead to nonlinear effects already at low driving forces. At
surfaces, particles can roll off, facilitating slip. More generally,
the interplay of rolling and sliding friction may give rise to
interesting structures at high shear rates, which we will explore in
future work.

In the case of MIPS, we have recovered the result of a previous
study\cite{nie2020frictional} that frictional contacts increase the
width of the phase-separated region. Beyond that, we have shown that
thermodynamic inconsistencies can lead to a wrong prediction of the
resulting collective behavior, in parameter regimes near the
transition line between the uniform and the MIPS phase. We anticipate
that the coupling of rotational and translational noise might be
relevant in active chiral matter and unveil new mechanisms of pattern
formation.

In future work, it will be interesting to extend our approach to  the
overdamped case. The present study suggests that also in underdamped
models, e.g., for active Brownian particles, the coupling of
rotational and translational stochastic terms will be crucial to
achieve thermodynamic consistency. Unfortunately, simply taking the
limit $m \to 0$ and $\II \to 0$ is tricky in the presence of
frictional contacts and will require special efforts.  

Furthermore, it will be interesting to examine the effect of phenomena
that have been neglected in the present work. In our model, we have
not included spinning friction, i.e., friction that occurs if two
particle faces are rotating against each other. 
\rev{This type of friction is not present in strictly two-dimensional
systems,} but may become important in three-dimensional or quasi
two-dimensional systems of translationally nearly immobile spinning
particles.  Applying our Fokker-Planck approach to deriving the
correct noise structure for this type of friction should be a
relatively straightforward exercise. We have also neglected memory
effects\cite{elperin1997comparing, herrmann1998modeling}, which become
important in systems that not only exhibit dynamic friction, but also
static friction. Modeling thermal fluctuations consistently in such
cases will be an interesting challenge for future work.

\section{Methods}

\subsection{Theory: Relation between Langevin equation and the Fokker-Planck equation}
\label{sec:A}
We recall the general relation between
the Langevin equation and the Fokker-Planck equation for systems with
many variables \cite{paulbaschnagel_book}. We consider a system of stochastic
differential equations for a vector of $N$ variables, 
$\bm{x} = \{x_i \}$ with the Langevin equation
\begin{equation}
\label{eq:LE_multivariate}
\dot{\bm{x}} = \bm{\mu}(\bm{x},t) + \bm{\sigma}(\bm{x},t) \: \bm{\eta}(t),
\end{equation}
where $\bm{\eta}(t)$ is an uncorrelated Gaussian white noise vector with
fluctuations
\begin{equation}
\label{eq:noise_multivariate}
\langle \eta_\alpha(t) \eta_\beta(t') \rangle
= \delta_{\alpha \beta} \: \delta(t-t')\, .
\end{equation} 
Then the corresponding Fokker-Planck equation for the
distribution function $P(\bm{x},t)$ is
\begin{equation}
\label{eq:FP_multivariate}
\partial_t {\cal P}
  = - \sum_\alpha \partial_\alpha \: \mu_\alpha(\bm{x}) \: {\cal P}
    - \sum_\alpha J_\alpha^{\text{R}} {\cal P}
\end{equation}
with \cite{paulbaschnagel_book,Sokolov2010ito}
\begin{equation}
\label{eq:FP_multivariate_Ito}
J_\alpha^{\text{R}} = -
   \frac{1}{2} \sum_{\beta,\gamma} \partial_\beta \:
                \sigma_{\alpha \gamma}(\bm{x},t) \sigma_{\beta \gamma}(\bm{x},t) 
\end{equation}
in the It\^o calculus,
\begin{equation}
\label{eq:FP_multivariate_Stratonovich}
J_\alpha^{\text{R}} =   
  - \frac{1}{2} \sum_{\beta,\gamma} \sigma_{\alpha \gamma}(\bm{x},t) \: 
     \partial_\beta  \: \sigma_{\beta \gamma}(\bm{x},t) 
\end{equation}
in the Stratonovich calculus, and 
\begin{equation}
\label{eq:FP_multivariate_HK}
J_\alpha^{\text{R}} =   
  - \frac{1}{2} \sum_{\beta,\gamma} \sigma_{\alpha \gamma}(\bm{x},t) \: 
     \sigma_{\beta \gamma}(\bm{x},t) \: \partial_\beta
\end{equation}
in the H\"anggi-Klimontovich calculus.
If $\bm{\sigma}$ does not depend on $\bm{x}$ (additive noise), then 
all formalisms give the same Fokker-Planck equation. 

With these general relations in mind, our task is to translate the 
Fokker-Planck probability operators ${\bf J}_{ik,\pp}^\text{R}$ and
${\bf J}_{ik,\jj}^\text{R}$, Eqs.\ (\ref{eq:FP_prob_pp}) and
(\ref{eq:FP_prob_jj}), into stochastic force and torque terms
in the equations of motion of the particles $i$ and $k$. 
In the notation of Eqs.\ (\ref{eq:LE_multivariate})
- (\ref{eq:FP_multivariate_HK}), the sum 
$ [{\bf J}_{ik,\pp}^{\text{R}} + {\bf J}_{ik,\jj}^{\text{R}}]_\alpha$
 corresponds to terms
 $-\frac{1}{2} \sum_{\beta,\gamma} 
  \partial_\beta \sigma^{(ik)}_{\alpha \gamma} 
   \sigma^{(ik)}_{\beta \gamma}$,
 $-\frac{1}{2} \sum_{\beta,\gamma} 
   \sigma^{(ik)}_{\alpha \gamma} \partial_\beta
   \sigma^{(ik)}_{\beta \gamma}$,
or
 $-\frac{1}{2} \sum_{\beta,\gamma} 
   \sigma^{(ik)}_{\alpha \gamma} 
   \sigma^{(ik)}_{\beta \gamma} \partial_\beta$,
depending on the calculus, where the indices $\alpha,\beta,\gamma$
can take 12 values corresponding to the components of the variables
$\pp_i, \pp_k, \jj_i, \jj_k$. We can decompose the matrices 
$\bm{\sigma}^{(ik)}$ as $\bm{\sigma}^{(ik)} = 
\sqrt{\kbT} D(u_{ik}^\perp,r_{ik}) \: \hat{\bm{\sigma}}^{(ik)}$ 
with reduced matrices $\hat{\bm{\sigma}}(\hr_{ik})$ that are 
independent of $\pp_i, \pp_k, \jj_i, \jj_k$ and satisfy
 \begin{equation*}
 \frac{1}{2} \: \hat{\bm{\sigma}} \hat{\bm{\sigma}}^T
 = 
 \mbox{\footnotesize $
 \left(
 \begin{array}{cc|cc}
 \PP(\hr_{ik}) & - \PP(\hr_{ik}) & 
 - R_i \bm{A}(\hr_{ik})  & - R_k \bm{A}(\hr_{ik}) \\
 - \PP(\hr_{ik}) &  \PP(\hr_{ik}) & 
  R_i \bm{A}(\hr_{ik})  &  R_k \bm{A}(\hr_{ik})  \\
 \hline
 R_i \bm{A}(\hr_{ik}) & - R_i \bm{A}(\hr_{ik}) &
 R_i^2 \PP(\hr_{ik}) & R_i R_k \PP(\hr_{ik}) \\
 R_k \bm{A}(\hr_{ik}) & - R_k \bm{A}(\hr_{ik}) &
 R_i R_k \PP(\hr_{ik}) & R_k^2 \PP(\hr_{ik}) 
 \end{array}
 \right)
 $}
 .
 \end{equation*}
 Here $\bm{A}(\hr_{ik})$ is defined as the antisymmetric matrix
 with $\bm{A}(\hr_{ik}) \bm{a} = \hr_{ij} \times \bm{a}$ for all
 $\bm{a}$. One possible solution for $\hat{\bm{\sigma}}$ is
 \begin{equation*}
 \hat{\bm{\sigma}}
 = 
 \mbox{\footnotesize $
 \left(
 \begin{array}{cc|cc}
 \PP(\hr_{ik}) & 0 & - \bm{A}(\hr_{ik})  & 0 \\
 - \PP(\hr_{ik}) & 0 & \bm{A}(\hr_{ik})  & 0 \\
 \hline
 R_i \bm{A}(\hr_{ik}) & 0 & R_i \PP(\hr_{ik}) & 0 \\
 R_k \bm{A}(\hr_{ik}) & 0 & R_k \PP(\hr_{ik}) & 0 
 \end{array}
 \right)
 $}
 .
 \end{equation*}
This can be checked using the relations
\begin{align*}
 \PP(\hr_{ik}) \PP(\hr_{ik})^T 
 = \bm{A}(\hr_{ik}) \bm{A}(\hr_{ik})^T
 & = \PP(\hr_{ik})
\\
 \bm{A}(\hr_{ik}) \PP(\hr_{ik})^T 
 = - \PP(\hr_{ik}) \bm{A}(\hr_{ik})^T
 &= \bm{A} (\hr_{ik}).
\end{align*}

With this choice of $\hat{\bm{\sigma}}$, half of the
 Gaussian random variables $\eta_i$ actually do not enter the Langevin
 equations of motion and can be omitted. We denote the other half as
 $\xx_{ij}^{\text{f}}=-\xx_{ji}^{\text{f}}$
 and $\NN_{ij}^{\text{f}}= \NN_{ji}^{\text{f}}$ and require
 them to be three-dimensional Gaussian white noise vectors with
 correlations
 \begin{eqnarray*}
 \langle \xx_{ij}^{\text{f}}(t) \xx_{kl}^{\text{f}}(t') \rangle
 &=& \kbT\:
    \one \: (\delta_{ik} \delta_{jl} - \delta_{il} \delta_{jk}) \:
   \delta(t-t')
 \\
 \langle \NN_{ij}^{\text{f}}(t) \NN_{kl}^{\text{f}}(t') \rangle
 &=& \kbT \:
    \one \: (\delta_{ik} \delta_{jl} + \delta_{il} \delta_{jk}) \:
   \delta(t-t').
 \end{eqnarray*}
 Then we can rewrite our random force and torque terms in the 
 form of Eqs.\ (\ref{eq:noise_force}) and (\ref{eq:noise_torque}). 

\subsection{Simulation parameters and units}
The natural time unit $t_0$ is given by the inertial time unit 
$t_0 = \sqrt{m \sigma^2/\kbT}$ in Sections \ref{sec:thermal} 
and \ref{sec:poiseuille}, and by the diffusive time unit $t_\text{R} = \gamma_R/\kbT$
in Section \ref{sec:mips}.  In Sections \ref{sec:thermal} and
\ref{sec:poiseuille}, we choose $\gamma_R = \pi \kbT t_0$, $\gamma =
\frac{3}{4 R^2} \gamma_R = 3 \pi \kbT t_0/\sigma^2$, corresponding to
the friction parameters of a spherical particle in a Newtonian fluid
with viscosity $\eta = 1 \: \kbT t_0/\sigma$ (i.e., $\gamma_R = 8 \pi
\eta R^3, \; \gamma = 6 \pi \eta R$).  In Section \ref{sec:mips}, we
choose the parameters to be the same as in a previous MIPS study
without frictional contacts\cite{Hecht2022}, i.e., $\gamma =
\gamma_R/\sigma^2 = t_\text{R} \kbT /\sigma^2$ and $m = 0.05 \:\kbT \:
(t_\text{R}/\sigma)^2$.  Furthermore, the parameter $\epsilon$ in the
WCA potential (Eq.\ (\ref{eq:wca})) is set to $\epsilon = \kbT$ in in
Sections \ref{sec:thermal} and \ref{sec:poiseuille}, and to $\epsilon
= 10 \kbT$ in Section \ref{sec:mips}.  The results presented in
Sections \ref{sec:thermal} and \ref{sec:poiseuille} were acquired
using HOOMD-blue \cite{HOOMD}, while those in Section \ref{sec:mips}
were obtained with LAMMPS \cite{LAMMPS}.

\subsection{Data availability}
The data that support the findings of this study are available from
the corresponding author upon reasonable request.

The simulation codes used to generate data (implemented in LAMMPS and
HOOMD-blue) are available at
\url{https://doi.org/10.48328/tudatalib-1838}. The HOOMD simulation
code has been merged with HOOMD-blue and is available in
the version 6.0.0, see
\url{https://hoomd-blue.readthedocs.io/en/v6.0.0/hoomd/md/pair/module-friction.html}
for a documentation.

\bibliography{refs, refsmemory}

\section{Acknowledgments}

We thank the anonymous reviewers for their careful assessment of the
manuscript and numerous helpful comments.  FS thanks Ralf Metzler for
pointing out the H\"anggi-Klimontovich formalism. This work was funded
by the Deutsche Forschungsgemeinschaft (DFG) via Grant 233630050, TRR
146, Project A3, and A9. The authors gratefully acknowledge the
computing time provided to them on the high-performance computer
Mogon2 and Mogon NHR South-West.

\section{Author contributions}

FS and BL designed the research and supervised the work. FS developed
the theory with support from KH and KD. KH and KD implemented the
simulation codes (HOOMD-blue and LAMMPS, respectively), performed the
simulations, analyzed and visualized the data. All authors contributed
to the writing of the original draft. KH and FS reviewed and finalized
the manuscript.  


\section{Competing interests}

The authors declare no competing interests.

\clearpage

\section{Supplementary information}
\subsection{Deterministic two particle collision}
\begin{figure}[H]
    \centering
    \includegraphics[width=0.8\linewidth]{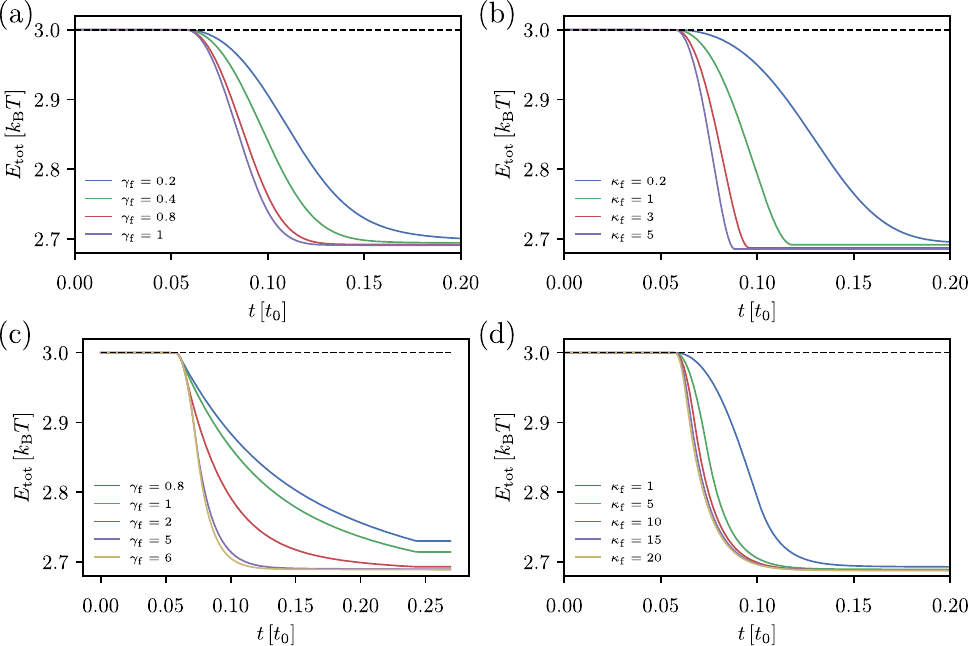}
    \caption{$E_\mathrm{tot}$ total energy over time $t$ for linear friction model (a) Coulomb friction model (b) and Coulomb-Newton friction model  with (c) fixed $\kappa_\mathrm{f}=5$ and varying $\gamma_\mathrm{f}$, and (d) fixed $\gamma_\mathrm{f}=6\: \kbT t_0/\sigma^2$ and varying $\kappa_\mathrm{f}$.}
    \label{fig:collision}
\end{figure}

In section II A 2, we analytically demonstrate that the deterministic friction force always reduces the kinetic energy of the particles during a collision. To validate this result, we simulated a collision between two particles in the absence of stochastic contact forces. Both particles have the radius $R=\sigma/2$ and mass $m=1\,k_\mathrm{B}T(t_0/\sigma)^2$. The first particle is initially placed at the origin, while the second is positioned at $(1.15,0.4,0)\,\sigma$. The first particle has an initial velocity $v_x=\sqrt{3}\,\sigma/t_0$ in the $x-$direction and an initial angular velocity $\omega_z=2\cdot\sqrt{30}\,t_0^{-1}$ in $z-$direction, whereas the second particle is initially at rest. Both particles move freely and are not coupled to a thermostat. As shown in Fig.~\ref{fig:collision}, the total energy $E_\mathrm{tot}$ decreases during the collision for all friction types, confirming our analytical prediction.

\subsection{Validation of the frictional model}
\begin{figure}[H]
    \centering
    \includegraphics[width=0.8\linewidth]{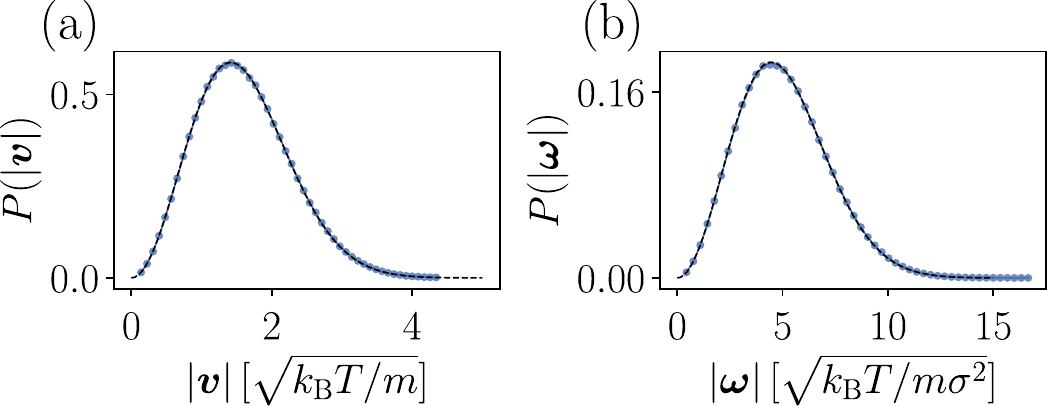}
    \caption{Speed (a) and angular speed (b) distribution in a
    three-dimensional system of passive colloids interacting with 
    repulsive WCA interactions and contact forces according to the  
    Coulomb-Newton friction model with noise, using $\gamma_{\text{f}} = 3 \cdot \sqrt{m \kbT}/\sigma$, $\kappa_{\text{f}} = 5$. Blue symbols show simulation results, lines show the Maxwell-Boltzmann distribution.}
    \label{fig:validation_CN}
\end{figure}

To confirm that our approach of frictional contacts yields a correct thermostat, we conducted a three-dimensional simulation of passive colloids interacting with repulsive WCA interactions and frictional contact interactions of the Coulomb-Newton type (with $\gamma_{\text{f}} = 3 \cdot \sqrt{m \kbT}/\sigma$ and $\kappa_{\text{f}} = 5$). The system contains $N=10000$ particles at a volume fraction of $\rho_\mathrm{volume}=0.15$. Periodic boundary conditions are applied in all three directions. The particles have the radius $R=\sigma/2$ and mass $m=1\,k_\mathrm{B}T(t_0/\sigma)^2$. The system is equilibrated over a time $2\,t_0$ using a time step of $\Delta t=0.001\,t_0$. The distribution of speed and angular speed in Fig.~\ref{fig:validation_CN} is then collected over a simulation time period of $4\,t_0$. They are in perfect agreement with the Maxwell-Boltzmann distribution, thus validating the accuracy of the method.

\subsection{Thermodynamic consistency in models with frictional contact interactions}
\begin{figure}
    \centering
    \includegraphics[width=0.8\linewidth]{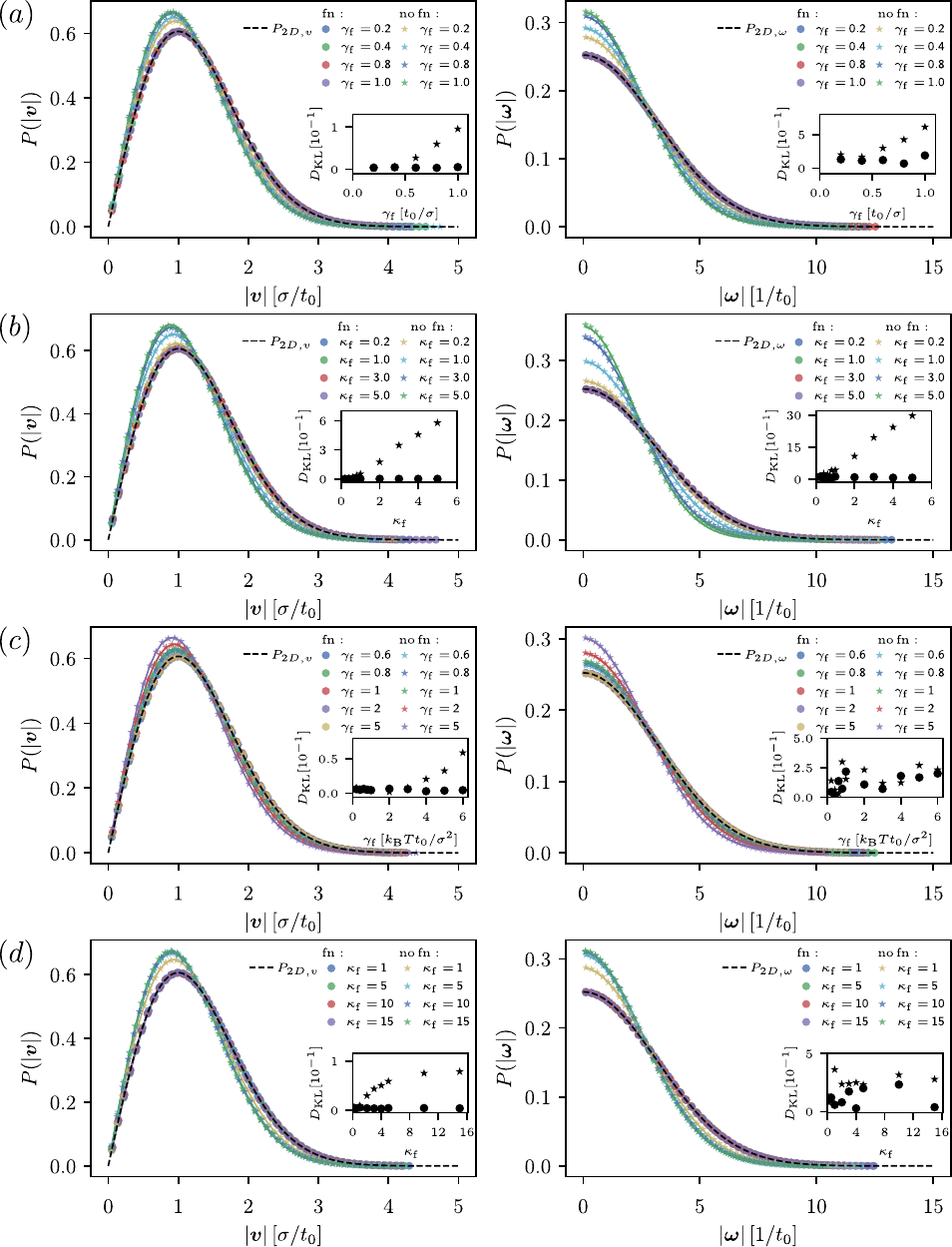} 
    \caption{(left) Speed histograms and (right) angular speed histograms for a two-dimensional passive colloid system with frictional contacts according to the (a) linear friction model (b) Coulomb friction model (c,d) Coulomb-Newton friction model. For different friction constants $\gamma_\text{f}$ and $\kappa_\text{f}$ as indicated. Stars (labelled "no fn") show data from simulations without stochastic contact friction terms, circles (labelled "fn") data for the full model. Solid colored lines are fits to appropriate Maxwell-Boltzmann distribution with fitted effective temperature, $T_\text{eff}$, black solid line gives the theoretical expectation for $k_\text{B} T_\text{eff} = k_\text{B}T$.  The inset shows the corresponding Kullback-Leibler divergence between
    the distribution measured in simulations and the fitted Maxwell-Boltzmann distribution.}\label{fig:2D}
\end{figure}

\begin{figure}
    \centering
    \includegraphics[width=0.8\linewidth]{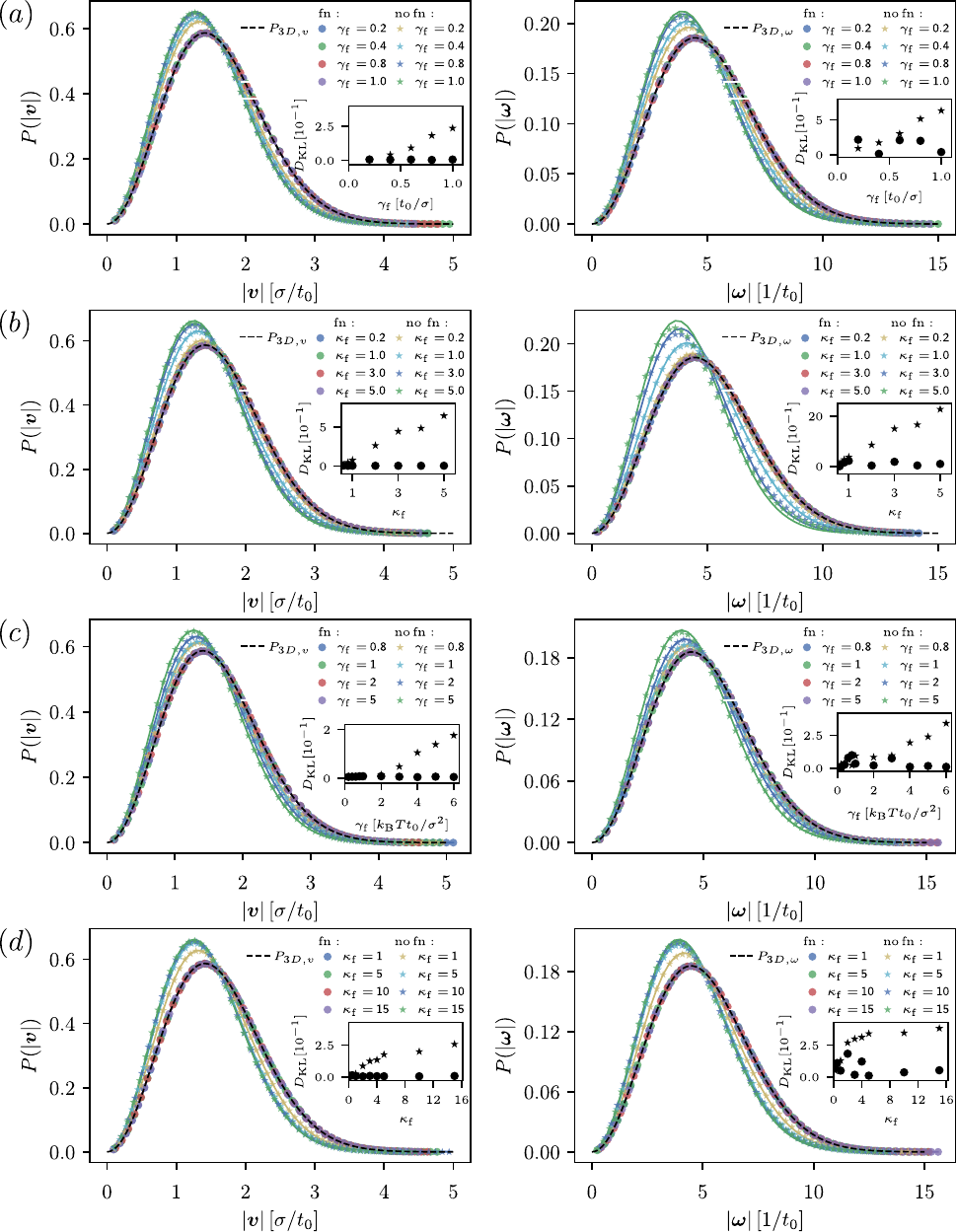}
    
    \caption{(left) Speed histograms and (right) angular speed histograms for a three-dimensional passive colloid system with frictional contacts according to the (a) linear friction model (b) Coulomb friction model (c,d) Coulomb-Newton friction model. For different friction constants $\gamma_\text{f}$ and $\kappa_\text{f}$ as indicated.  Stars (labelled "no fn") show data from simulations without stochastic contact friction terms, circles (labelled "fn") data for the full model. Solid colored lines are fits to the Maxwell-Boltzmann distributions with fitted effective temperature, $T_\text{eff}$, black solid line gives the theoretical expectation for $k_\text{B} T_\text{eff} = k_\text{B}T$.  The inset shows the corresponding Kullback-Leibler divergence between
    the distribution measured in simulations and the fitted Maxwell-Boltzmann distribution.}\label{fig:3D}
\end{figure}

In section II B 1, we discuss the speed and angular speed histogram for a two-dimensional system with frictional contacts according to the Coulomb friction model, which are used to determine the effective temperatures $T_\mathrm{eff}$. Here we additionally show the corresponding histograms for both two-dimensional (Figure \ref{fig:2D}) and three-dimensional (Figure \ref{fig:3D}) systems with the linear friction model, the Coulomb friction model, and the Coulomb-Newton friction model. The insets show the corresponding Kullback-Leibler divergence between the distribution obtained in simulations and the fitted Maxwell-Boltzmann distribution. As discussed in the main manuscript,the Kullback-Leibler divergences increase with the friction parameters in the absence of frictional noise, indicating that the shape of the distribution from the simulations generally deviates from the fitted Maxwell-Boltzmann distribution for the majority of systems. One notable exception is the angular speed distribution in the two-dimensional Coulomb-Newton case (Fig.~\ref{fig:2D} c) and d)), where we did not find a significant influence of the frictional noise on the Kullback-Leibler divergence, as the values lack a clear trend. We note that the Kullback-Leibler divergence is always positive by construction, therefore statistical spread necessarily results in a
positive value. However, taking the limit $\kappa_\text{f}\to0$ as a reference, where the frictional contact forces are essentially turned off, one can see that the values of the Kullback-Leibler divergences never substantially deviate from these reference values. We conclude
that for Coulomb-Newton friction with $\gamma_\text{f} < 6k_\text{B}T\,t_0/\sigma^2$, the use of an inconsistent thermostat does not seem to affect the functional form of the angular speed distribution $P(|\bm{\omega}|)$ in two-dimensions (within the error). It does however affect the form of the speed distribution $P(|\bm{v}|)$ and all effective temperatures.

\subsection{Stationary state of pressure-driven flows}
\begin{figure}[H]
    \centering
    \includegraphics[width=\linewidth]{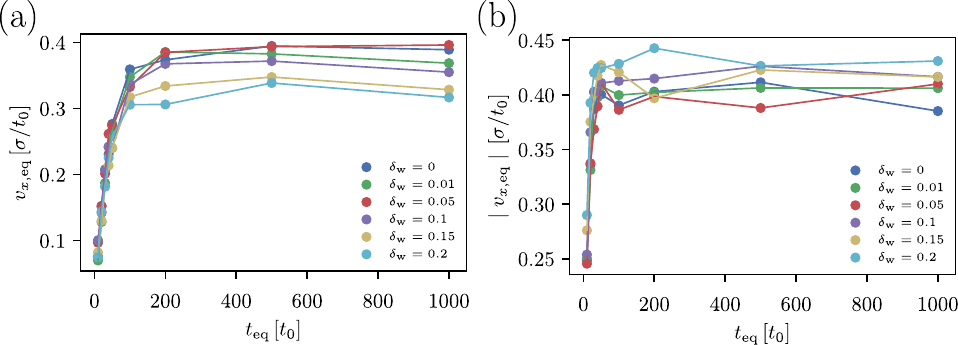} 
    \caption{(a) Mean particle velocity $v_{x,\mathrm{eq}}$ in the direction of the bulk force and and (b) absoulute mean particle velocity $|v_{x,\mathrm{eq}}|$ in the direction of the sliding walls} of particles for various wall roughness $\delta_\mathrm{w}$ after the time $t_\mathrm{eq}$.
    \label{fig:steadystate}
\end{figure}

In section II B 2, we examine the dynamics of repulsive frictional particles under an external bulk force and sliding walls in a slit channel. Figure~\ref{fig:steadystate} a) displays the average particle velocity $v_{x,\mathrm{eq}}$ along the direction of the bulk force for varying wall roughness parameters $\delta_\mathrm{w}$, measured after a simulation time $t_\mathrm{eq}$. Figure~\ref{fig:steadystate} b) shows the averaged particle velocity $|v_{x,\mathrm{eq}}|$, where we determined the mean velocity within each channel half ($y > 0$ and $y < 0$) and then averaged these two values, using the absolute value for the lower half. Our results show that both system reach a stationary state by $t_\mathrm{eq}=500\,t_0$ attaining a constant velocity, independent of further simulation time for each wall roughness parameters.

\subsection{Determination of the slip length and hydrodynamic boundary}
\begin{figure}[H]
    \centering
    \includegraphics[width=\linewidth]{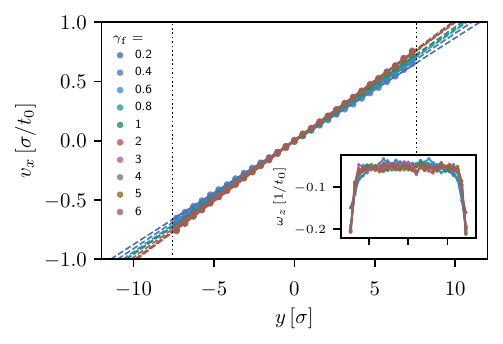} 
    \caption{Example of velocity profiles for fluids with frictional contact interactions of the Coulomb-Newton type with parameter $\kappa_\text{f}=5$ and different $\gamma_\text{f}$ as indicated, which are confined between oppositely sliding walls with wall velocity $v_\text{wall}=\pm1\,\sigma/t_0$ and wall roughness $\delta_\text{w}=0$. Solid lines show fit to a linear profile. Inset shows corresponding profiles of the $z$-component of the angular velocity.}
    \label{fig:Profile_couette}
\end{figure}

Section II B 2 presents the slip length and hydrodynamic boundary of a wall composed of frictional particles, which we obtained by analyzing the velocity profiles from Poiseuille and Couette flow simulations. The parabolic velocity profile of the Poiseuille flow simulations is shown in the main manuscript for varying friction parameter $\gamma_\text{f}$ and a wall roughness of $\delta_\text{w}=0$. Fig.~\ref{fig:Profile_couette} presents the velocity profile of the corresponding Couette flow simulations with a wall velocity of $v_{wall}=\pm1\,\sigma/t_0$.

\subsection{Influence of translational and rotational noise}
\begin{figure}[h]
    \centering
    \includegraphics[width=\linewidth]{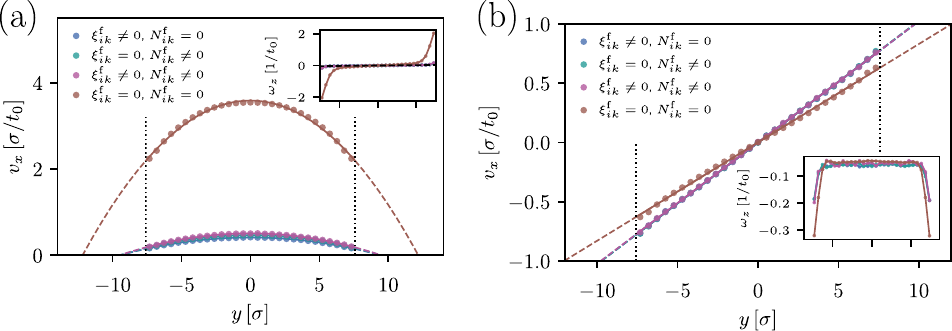} 
    \caption{Velocity profiles for fluids with frictional contact interactions of the Coulomb-Newton type with parameter $\kappa_\text{f}=5$ and $\gamma_\text{f}=6$, which are (a) confined between walls and subject to a bulk force $F=0.01\,k_\text{B}T/\sigma$ or (b) confined between oppositely sliding walls with wall velocity $v_\text{wall}=\pm1\,\sigma/t_0$ and wall roughness $\delta_\text{w}=0$. Solid lines show fit to an parabolic or linear profile. Insets show corresponding profiles of the $z$-component of the angular velocity.}
    \label{fig:Profile_poiseuille_couette_noise}
\end{figure}

 To isolate the effect of the translational and rotational frictional noise on the fluid, we performed Poiseuille and Couette flow simulations with these noises selectively disabled. Translation frictional noise is deactivated by setting the random force $\bm{F}^\text{R}_{ik}=0$ and the term $\xi^\text{f}_{ik}=0$ (see Eq.~(24) and (25) of the main manuscript). Analogously, rotational noise is removed by setting the random torque $\bm{\tau}^\text{R}_{ik}=0$ and the term $N^\text{f}_{ik}=0$. Fig.~\ref{fig:Profile_poiseuille_couette_noise} shows the resulting velocity profiles for (a) Poiseuille flow and (b) Couette flow. The inset shows the corresponding profiles of the $z$-component of the angular velocity. Compared to the thermodynamically consistent case, deactivating just one type of frictional noise does not alter the velocity profiles in Poiseuille or Couette flow. Interestingly, this is different if frictional noise is completely disabled, which corresponds to setting the temperature to zero. For Poiseuille flow, where particles are driven by a constant bulk force, omitting frictional noise entirely significantly increases their velocity as well as the $z$-component of the angular velocity near the walls. In contrast, for Couette flow, in which particles are driven by the frictional interaction with the oppositely moving walls, omitting frictional noise entirely decreases the absolute velocity. Fitting the Poiseuille flow profiles to a parabola, we find that the the viscosity drops from $\eta = (0.254 \pm  \: 0.001) \epsilon t_0/\sigma$ at temperature $k_\text{B}T = 1 \: \epsilon$ to $\eta = ( 0.059 \pm 0.001)  \: \epsilon t_0/\sigma$ at temperature $k_\text{B}T=0$. Thus, frictional contact forces are activated by temperature. An intuitive explanation is particles that in the athermal flows  almost stop moving perpendicular to the bulk force or sliding walls. This results in fewer particle collisions and thus in fewer frictional contacts.
\end{document}